\documentclass[journal]{IEEEtran}

\input{Qcircuit.tex }
\usepackage{graphicx} 
\usepackage{amsmath} 
\usepackage{amssymb}
\usepackage{amsthm}
\usepackage{subcaption}
\usepackage{float}
\usepackage{epstopdf}
\usepackage{multicol}
\usepackage{multirow}
\usepackage{booktabs}
\usepackage[autostyle=false, style=english]{csquotes}
\MakeOuterQuote{"}
\usepackage{color}
\ifCLASSINFOpdf
\else
\fi

\begin{document}
%
% paper title
% Titles are generally capitalized except for words such as a, an, and, as,
% at, but, by, for, in, nor, of, on, or, the, to and up, which are usually
% not capitalized unless they are the first or last word of the title.
% Linebreaks \\ can be used within to get better formatting as desired.
% Do not put math or special symbols in the title.
\title{T-count and Qubit Optimized Quantum Circuit Designs of Carry Lookahead Adder}

% author names and affiliations
% use a multiple column layout for up to three different
% affiliations
\author{Himanshu Thapliyal, Edgard Mu\~{n}oz-Coreas, Vladislav Khalus
\thanks{ }
\thanks{Himanshu Thapliyal, Edgard Mu\~{n}oz-Coreas and Vladislav Khalus are with the Department of Electrical and Computer Engineering, University of Kentucky, Lexington, KY, USA. \newline
Email: hthapliyal@uky.edu}}

%\bstctlcite{IEEEexample:BSTcontrol}
% conference papers do not typically use \thanks and this command
% is locked out in conference mode. If really needed, such as for
% the acknowledgment of grants, issue a \IEEEoverridecommandlockouts
% after \documentclass

% for over three affiliations, or if they all won't fit within the width
% of the page, use this alternative format:
% 
%\author{\IEEEauthorblockN{Michael Shell\IEEEauthorrefmark{1},
%Homer Simpson\IEEEauthorrefmark{2},
%James Kirk\IEEEauthorrefmark{3}, 
%Montgomery Scott\IEEEauthorrefmark{3} and
%Eldon Tyrell\IEEEauthorrefmark{4}}
%\IEEEauthorblockA{\IEEEauthorrefmark{1}School of Electrical and Computer Engineering\\
%Georgia Institute of Technology,
%Atlanta, Georgia 30332--0250\\ Email: see http://www.michaelshell.org/contact.html}
%\IEEEauthorblockA{\IEEEauthorrefmark{2}Twentieth Century Fox, Springfield, USA\\
%Email: homer@thesimpsons.com}
%\IEEEauthorblockA{\IEEEauthorrefmark{3}Starfleet Academy, San Francisco, California 96678-2391\\
%Telephone: (800) 555--1212, Fax: (888) 555--1212}
%\IEEEauthorblockA{\IEEEauthorrefmark{4}Tyrell Inc., 123 Replicant Street, Los Angeles, California 90210--4321}}

% use for special paper notices
%\IEEEspecialpapernotice{(Invited Paper)}

% make the title area
\maketitle

% As a general rule, do not put math, special symbols or citations
% in the abstract
\begin{abstract}
	
	Quantum circuits of arithmetic operations such as addition are needed to implement quantum algorithms in hardware.  Quantum circuits based on Clifford+T gates are used as they can be made tolerant to noise.  The tradeoff of gaining fault tolerance from using Clifford+T gates and error correcting codes is the high implementation overhead of the T gate.  As a result, the T-count performance measure has become important in quantum circuit design.  Due to noise, the risk for errors in a quantum circuit computation increases as the number of gate layers (or depth) in the circuit increases.  As a result, low depth circuits such as quantum carry lookahead adders (QCLA)s have caught the attention of researchers. This work presents two QCLA designs each optimized with emphasis on T-count or qubit cost respectively.    In-place and out-of-place versions of each design are shown. The proposed QCLAs are compared against the existing works in terms of T-count.  The proposed QCLAs for out-of-place addition achieve average T gate savings of $54.34 \%$ and $37.21 \%$, respectively.  The proposed QCLAs for in-place addition achieve average T gate savings of $72.11 \%$ and $35.87 \%$ respectively.          
	
\end{abstract}

% no keywords

% for peer review papers, you can put extra information on the cover
% page as needed:
% \ifCLASSOPTIONpeerreview
% \begin{center} \bfseries EDICS Category: 3-BBND \end{center}
% \fi
%
% For peerreview papers, this IEEEtran command inserts a page break and
% creates the second title. It will be ignored for other modes.
\IEEEpeerreviewmaketitle

\section{Introduction}

Quantum circuits of arithmetic operations are needed to design quantum hardware for implementing quantum algorithms such 
As integer factoring, searching and quantum mechanical simulation.  Quantum circuits for addition are fundamental building 
blocks crucial to implementing these quantum algorithms \cite{Childs2012Hamiltonian} \cite{Shor1994factoring-discretelog} 
\cite{Novo2017Hamiltonian} \cite{Hallgren2007quadraticmotivation} \cite{Childs2013discretelog}.  Thus, researchers have 
invested considerable effort in designing quantum adders such as ripple carry adders or carry lookahead adders (QCLAs) 
\cite{Lidia2017adder} \cite{Draper2004Adder} \cite{Takahashi2008qcla} \cite{Munoz2019adder} \cite{Quipper} \cite{LIQUi}.   

Quantum circuits possess properties that make them distinct from circuits for other technologies.  For example, there is a one-to-one relationship between the inputs and outputs in a quantum circuit.  As a result, the quantum circuit designer will 
face additional sources of circuit overhead such as ancillae and garbage output.  Ancillae are constant inputs to the 
quantum circuit.  Garbage output are any circuit output that is not a circuit input or needed output.  To make full use of 
the qubit resources of the quantum machine, the garbage output will need to be cleared.  This process will add to the 
overall qubit cost and gate cost of a quantum circuit and is discussed in \cite{Bennett1973trashremoval}.

Physical quantum computers are prone to noise errors \cite{Kiteav2005faulttolerant} \cite{Mosca2} \cite{PalerIOP}.  
Quantum circuits based on Clifford+T gates have caught the attention of researchers because can be made fault tolerant 
with quantum error correcting codes permitting reliable and scalable quantum computation \cite{PalerIOP} \cite{Zhou-T}.  
This set is universal and has been used to realize basic reversible logic gates and larger functional blocks \cite{Gosset} 
\cite{Maslov} \cite{Zhou-T} \cite{Cody2012toffoligatebuild} \cite{Miller}.  However, fault tolerance comes with increased 
implementation overhead especially the overhead associated with implementing the T gate \cite{Zhou-T} \cite
{Devitt2013Tgatebuild} \cite{Fowler2012faulttolerance}.  The high cost to implement the T gate has caused the measures of 
T-count and T-depth to become important measures to evaluate quantum circuit cost \cite{Gosset} \cite{Mosca2} \cite
{Gidney20184TgateToffolibuild} \cite{Munoz2019adder}.

QCLAs have caught the interest of researchers because they perform the addition operation in order $\mathcal{O}(log(n))$ 
circuit depth (while the ripple carry adder has a depth of $\mathcal{O}(n)$).  Low-depth circuits such as QCLAs have use 
in quantum hardware applications where longer computation time increases the risk of errors \cite{Meter2006coherencetimes} 
\cite{IBM_quantum} \cite{Yu2016coherencetime}.  Several QCLA designs have been proposed in the literature such as \cite
{Cheng2002QCLAinplace} or \cite{Takahashi2008qcla}.  Designs that return the sum on ancillae (or out-of-place QCLA) and 
designs that replace one of the primary inputs with the sum (or in-place QCLA) have been proposed.  Table \ref{Comparison-of-OOP-works} and Table \ref{Comparison-of-IP-works} summarizes the existing work.  Table \ref{Comparison-of-OOP-works} presents existing out-of-place QCLAs and Table \ref{Comparison-of-IP-works} presents in-place QCLAs.  Further discussion on the existing work is in Section \ref{prior-works-qcla}.

\begin{table}[hbtp]
	\centering
	\scriptsize	
	\caption{Details of Existing Out-of-Place QCLAs}
	\label{Comparison-of-OOP-works}
	\begin{tabular}{p{.5in}cccp{.575in}c}
		\\ \midrule
		\multicolumn{1}{c}{Design} & & T-count & qubits & gates used & garbage? \\ \cmidrule{1-1} \cmidrule{3-6}   
		Draper et al. (\cite{Draper2006qcla}) & & $\mathcal{O}(35 \cdot n)$ &  $\mathcal{O}(4 \cdot n)$ &  CNOT, Toffoli & no \\
		Babu et al.* (\cite{Babu2013QCLAoutofplace}) & & $\mathcal{O}(54 \cdot n)$ &  $\mathcal{O}(12 \cdot n)$ & PFA$^1$, CNOT, MIG$^2$, NFT$^3$ & yes \\
		Trisetyarso et al. (\cite{Trisetyarso2009qcla}) & & $\mathcal{O}(35 \cdot n)$ &  $\mathcal{O}(4 \cdot n)$ & NOT, CNOT, Toffoli, Hadamard, DCZ$^4$ & no \\
		Lisa et al.* (\cite{Lisa2015QCLAoutofplace}) & & $\mathcal{O}26 \cdot n)$ &  $\mathcal{O}(6 \cdot n)$ & CNOT, RPA$^5$, Fredkin$^7$ & yes \\
		Thapliyal et al. (\cite{Thapliyal2013qcla}) & & $\mathcal{O}(35 \cdot n)$ &  $\mathcal{O}(4 \cdot n)$ & CNOT,  Toffoli, Peres$^6$ & no \\ \bottomrule  
		\multicolumn{6}{p{3in}}{$^1$ The PFA gate decomposes into two V gates, a V+ gate and six CNOT gates.  Each V gate and V+ gate has a T-count of 3.  \cite{Babu2013QCLAoutofplace} \cite{Jamal2013gatebreakdown} \cite{Maslov}} \\
		\multicolumn{6}{p{3in}}{$^2$ The MIG gate decomposes into two V gates, a V+ gate and four CNOT gates. \cite{Jamal2013gatebreakdown}} \\
		\multicolumn{6}{p{3in}}{$^3$ The NFT gate decomposes into two V gates, a V+ gate and seven CNOT gates. \cite{Mitra2012gatebreakdowns}} \\
		\multicolumn{6}{p{3in}}{$^4$ The DCZ or double controlled Z gate has a T-count of seven. \cite{selinger2013gatebreakdown}} \\
		\multicolumn{6}{p{3in}}{$^5$ The RPA gate decomposes into a V gate, a V+ gate and three CNOT gates. \cite{Lisa2015QCLAoutofplace}} \\
		\multicolumn{6}{p{3in}}{$^6$ The Peres gate has a T-count of seven. \cite{Maslov}} \\
		\multicolumn{6}{p{3in}}{$^7$ The Fredkin gate has a T-count of seven. \cite{Maslov}} \\
		\multicolumn{6}{p{3in}}{* T-count and qubit cost are for the circuit after being modified to remove garbage output. We use the methodology outlined in \cite{Bennett1973trashremoval} to remove the garbage outputs}  \\
	\end{tabular}
\end{table}

\begin{table}[hbtp]
	\centering
	\scriptsize
	\caption{Details of Existing In-Place QCLAs}
	\label{Comparison-of-IP-works}
	\begin{tabular}{p{.475in}cccp{.575in}c}
		\\ \midrule
		\multicolumn{1}{c}{Design} & & T-count & qubits & gates used & garbage?  \\ \cmidrule{1-1} \cmidrule{3-6} 
		Draper et al. (\cite{Draper2006qcla}) & & $\mathcal{O}(70 \cdot n)$ & $\mathcal{O}(4 \cdot n)$ & NOT, CNOT, Toffoli & no \\ 
		Trisetyarso et al. (\cite{Trisetyarso2009qcla}) & & $\mathcal{O}(70 \cdot n) $ & $\mathcal{O}(4 \cdot n)$ & NOT, CNOT, Toffoli, Hadamard, DCZ$^1$ & no \\
		Thapliyal et al. (\cite{Thapliyal2013qcla}) & & $\approx \mathcal{O}(51 \cdot n)$ & $\mathcal{O}(4 \cdot n)$ & NOT, CNOT, Toffoli, Peres$^2$, TR$^3$ & no \\   
		Takahashi et al.(\cite{Takahashi2008qcla}) & & $\approx \mathcal{O}(196 \cdot n)$ & $\approx \mathcal{O}(5 \cdot n)$ & NOT, CNOT, Toffoli & no \\
		Takahashi et al.(\cite{Takahashi2010inlaceQCLA}) & & $\approx \mathcal{O}(49 \cdot n)$ & $\approx \mathcal{O}(5 \cdot n)$ &  NOT, CNOT, Toffoli & no \\
		Cheng et al. (\cite{Cheng2002QCLAinplace}) & & $\mathcal{O}(n^3)$ & $\mathcal{O}(3 \cdot n)$ & CNOT, Toffoli, MCT$^4$ & no \\
		Mogenson 1 (\cite{Mogensen2019qcla}) & & $\approx \mathcal{O}(42 \cdot n)$ & $\mathcal{O}(3 \cdot n)$ & CNOT, Toffoli, Fredkin$^5$ & no \\
		Mogenson 2 (\cite{Mogensen2019qcla}) & & $\approx \mathcal{O}(42 \cdot n)$ & $\mathcal{O}(3 \cdot n)$ & CNOT, Toffoli, Fredkin$^5$ & no \\ \bottomrule  
		\multicolumn{6}{p{3in}}{$^1$ The DCZ or double controlled Z gate has a T-count of seven. \cite{selinger2013gatebreakdown}} \\
		\multicolumn{6}{p{3in}}{$^2$ The Peres gate has a T-count of seven. \cite{Maslov}} \\
		\multicolumn{6}{p{3in}}{$^3$ The TR gate decomposes into two V gates, a V+ gate and a CNOT gate.  Each V gate and V+ gate has a T-count of 3. \cite{Boss2013gatebreakdown} \cite{Maslov}} \\
		\multicolumn{6}{p{3in}}{$^4$ The MCT or Multiple-Control Toffoli gate decomposes into $2 \cdot n - 3$ Toffoli gates.  The T-count is $14 \cdot n - 21$. } \\
		\multicolumn{6}{p{3in}}{$^5$ The Fredkin gate has a T-count of seven. \cite{Maslov}} \\
	\end{tabular}
	
\end{table}

\textit{To overcome shortcomings in existing works, this work proposes a family of designs for QCLAs.  The first design (FT-QCLA1) is optimized with an emphasis on T-count.  The second design (FT-QCLA2) is optimized with an emphasis on number of qubits.  All proposed designs enjoy reduced T gate cost and qubit cost compared to the existing works.  In-place QCLA implementations (In-FT-QCLA1 and In-FT-QCLA2) and out-of-place QCLA implementations (Out-FT-QCLA1 and Out-FT-QCLA2) for each design are shown.}  All proposed QCLAs can be made fault tolerant with error correcting codes.  The proposed QCLA designs are based on the NOT gate, CNOT gate, Toffoli gate, logical AND gate and uncomputation gate.  The logical AND gate and uncomputation gate are presented in \cite{Cody2012toffoligatebuild} and \cite{Gidney20184TgateToffolibuild}.  The 
proposed QCLAs are compared against existing works and shown to be superior in terms of qubit cost and T-count.         

This work is organized as follows:  Section \ref{QCLAs-background} provides background on the Clifford+T quantum gate set, 
introduces the logical AND gate and the uncomputation gate.  Section \ref{OOP-QCLA-design} presents the proposed out-of-place QCLAs.  Each design (Out-FT-QCLA1 and Out-FT-QCLA2) is addressed in its own subsection within Section \ref{OOP-QCLA-design}.  Section \ref{OOP-QCLA-performance} illustrates the comparison of the proposed out-of-place QCLAs against existing out-of-place QCLAs.  Section \ref{IP-QCLA-design} presents the proposed in-
place QCLAs.  Each design (In-FT-QCLA1 and In-FT-QCLA2) is addressed in its own subsection within Section \ref{IP-QCLA-design}.  Section \ref{IP-QCLA-performance} illustrates the comparison of the proposed in-place QCLAs against existing in-place QCLAs.  Section \ref{prior-works-qcla} provides a review of the existing work in quantum carry lookahead addition.

\section{Background}
\label{QCLAs-background}

\subsection{Quantum Gates}

\begin{figure}[hbt]
	\centering
	\begin{subfigure}[hb]{2.5in}
		\begin{center}
			\textsc{Clifford+T Gate Set}
		\end{center}
	\end{subfigure} \\ \begin{subfigure}[hb]{1in}
		Hadamard Gate
	\end{subfigure} \qquad \begin{subfigure}[hb]{.5in}
		\[
		\Qcircuit @C=0.7em @R=0.5em @!R{
			& \gate{H} &  \qw & &   }
		\]
	\end{subfigure} \qquad \begin{subfigure}[hb]{.75in}
		\centering
		$ \frac{1}{\sqrt{2}}
		\begin{bmatrix}
		1 & 1  \\
		1 & -1 
		\end{bmatrix} $
	\end{subfigure} \\ \begin{subfigure}[hb]{1in}
		
	\end{subfigure} \qquad \begin{subfigure}[hb]{.5in}
		
	\end{subfigure} \qquad \begin{subfigure}[hb]{.75in}
		
	\end{subfigure} \\ \begin{subfigure}[hb]{1in}
		T Gate
	\end{subfigure} \qquad \begin{subfigure}[hb]{.5in}
		\[
		\Qcircuit @C=0.7em @R=0.5em @!R{
			& \gate{T} &  \qw & &   }
		\]
	\end{subfigure} \qquad \begin{subfigure}[hb]{.75in}
		\centering
		$\begin{bmatrix}
		1 & 0  \\
		0 & e^{i \cdot \frac{\pi}{4}} 
		\end{bmatrix} $
	\end{subfigure} \\ \begin{subfigure}[hb]{1in}
		
	\end{subfigure} \qquad \begin{subfigure}[hb]{.5in}
		
	\end{subfigure} \qquad \begin{subfigure}[hb]{.75in}
		
	\end{subfigure} \\ \begin{subfigure}[hb]{1in}
		\flushleft
		Hermitian of T Gate
	\end{subfigure} \qquad \begin{subfigure}[hb]{.5in}
		\[
		\Qcircuit @C=0.7em @R=0.5em @!R{
			& \gate{T^{\dag}} &  \qw & &   }
		\]
	\end{subfigure} \qquad \begin{subfigure}[hb]{.75in}
		\centering
		$\begin{bmatrix}
		1 & 0  \\
		0 & e^{-i \cdot \frac{\pi}{4}} 
		\end{bmatrix} $
	\end{subfigure} \\ \begin{subfigure}[hb]{1in}
		
	\end{subfigure} \qquad \begin{subfigure}[hb]{.5in}
		
	\end{subfigure} \qquad \begin{subfigure}[hb]{.75in}
		
	\end{subfigure} \\ \begin{subfigure}[hb]{1in}
		Phase Gate
	\end{subfigure} \qquad \begin{subfigure}[hb]{.5in}
		\[
		\Qcircuit @C=0.7em @R=0.5em @!R{
			& \gate{S} &  \qw & &   }
		\]
	\end{subfigure} \qquad \begin{subfigure}[hb]{.75in}
		\centering
		$\begin{bmatrix}
		1 & 0  \\
		0 & i 
		\end{bmatrix} $
	\end{subfigure} \\ \begin{subfigure}[hb]{1in}
		
	\end{subfigure} \qquad \begin{subfigure}[hb]{.5in}
		
	\end{subfigure} \qquad \begin{subfigure}[hb]{.75in}
		
	\end{subfigure} \\ \begin{subfigure}[hb]{1in}
		\flushleft
		Hermitian of Phase Gate
	\end{subfigure} \qquad \begin{subfigure}[hb]{.5in}
		\[
		\Qcircuit @C=0.7em @R=0.5em @!R{
			& \gate{S^{\dag}} &  \qw & &   }
		\]
	\end{subfigure} \qquad \begin{subfigure}[hb]{.75in}
		\centering
		$\begin{bmatrix}
		1 & 0  \\
		0 & -i 
		\end{bmatrix} $
	\end{subfigure} \\ \begin{subfigure}[hb]{1in}
		
	\end{subfigure} \qquad \begin{subfigure}[hb]{.5in}
		
	\end{subfigure} \qquad \begin{subfigure}[hb]{.75in}
		
	\end{subfigure} \\ \begin{subfigure}[hb]{1in}
		NOT Gate
	\end{subfigure} \qquad \begin{subfigure}[hb]{.5in}
		\[
		\Qcircuit @C=0.7em @R=0.5em @!R{
			& \targ &  \qw & &   }
		\]
	\end{subfigure} \qquad \begin{subfigure}[hb]{.75in}
		\centering
		$\begin{bmatrix}
		0 & 1  \\
		1 & 0 
		\end{bmatrix}$
	\end{subfigure} \\ \begin{subfigure}[hb]{1in}
		
	\end{subfigure} \qquad \begin{subfigure}[hb]{.5in}
		
	\end{subfigure} \qquad \begin{subfigure}[hb]{.75in}
		
	\end{subfigure} \\ \begin{subfigure}[hb]{1in}
		Feynman (CNOT) Gate
	\end{subfigure} \qquad \begin{subfigure}[hb]{.5in}
		\[
		\Qcircuit @C=0.7em @R=0.5em @!R{
			& \ctrl{1} &  \qw & & \\ 
			& \targ &  \qw & & }
		\]
	\end{subfigure} \qquad \begin{subfigure}[hb]{.75in}
		\centering
		$\begin{bmatrix}
		1 & 0 & 0 & 0 \\
		0 & 1 & 0 & 0\\
		0 & 0 & 0 & 1 \\
		0 & 0 & 1 & 0 \\
		\end{bmatrix} $ 
	\end{subfigure}
	\caption{The quantum gate set used in this work.}
	\label{Clifford table}
\end{figure}

The proposed QCLA circuits in this work are based on the Clifford+T quantum gate set shown in Figure \ref{Clifford table}.  The Clifford+T gates have caught the interest of researchers because the gate set can be made tolerant to noise errors \cite{Kiteav2005faulttolerant} \cite{Mosca2} \cite{PalerIOP}.  The Clifford+T gate set is universal in nature permitting the fault tolerant quantum realization of any function of interest \cite{Boykin2000cliffordTgate} \cite{Gosset}.  However, the fault-tolerance comes with the trade-off of the high implementation costs of the T gate relative to the Clifford gates.  Details on the fault tolerant implementation of the T gate (and its associated costs) are in \cite{Devitt2013Tgatebuild} \cite{Fowler2012faulttolerance}.   The T gate is required because the Clifford gates by themselves are not universal \cite{Gosset} \cite{Mosca2} \cite{Boykin2000cliffordTgate} \cite{Zhou-T}.  The high implementation cost of the T gate has made T-count and T-depth metrics of interest to evaluate quantum circuit performance.  Existing quantum computers (such as those shown in \cite{IBM_quantum}) have a small number of qubits available.  As a result, the number of qubits in a quantum circuit is an important cost measure.  We now define the T-count, T-depth and qubit cost resource measures.

\begin{itemize} 
	\item T-count: T-count is the total number of T gates used in the quantum circuit.
	\item T-depth: T-depth is the number of T gate layers in the circuit, where a layer consists of quantum operations that can be performed simultaneously.
	\item Qubit cost: Qubit cost is the total number of qubits required to design the quantum circuit.
\end{itemize}

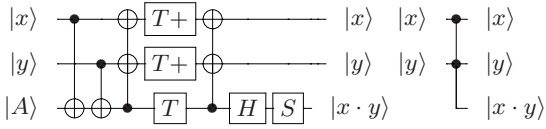
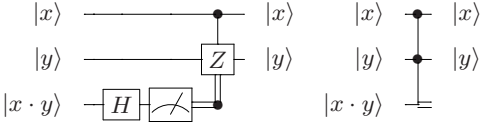
\begin{figure}
	\centering
	\begin{subfigure}[hb]{3in}
		\small
		\[
		\Qcircuit @C=0.3em @R=0.5em @!R{
			\lstick{\ket{x}} & & \ctrl{2} &\qw &\targ & \gate{T+} & \targ & \qw & \qw & \qw & \qw &\qw &\rstick{\ket{x}} & & & & & & & & & & & & & & & \lstick{\ket{x}} & & \ctrl{1} & \qw & \rstick{\ket{x}}\\
			\lstick{\ket{y}} & & \qw &\ctrl{1} &\targ & \gate{T+} & \targ & \qw & \qw & \qw & \qw &\qw &\rstick{\ket{y}}  & & & & & & & &   & & & & & & & \lstick{\ket{y}} & & \ctrl{1} & \qw & \rstick{\ket{y}} \\
			\lstick{\ket{A}} & & \targ &\targ &\ctrl{-2} & \gate{T} & \ctrl{-2} & \gate{H} & \gate{S} &\qw &\rstick{\ket{x \cdot y}} & & & & & & & & & & & & &  & & & & & & \qwx &\qw &\rstick{\ket{x \cdot y}} \\  
		}
		\]
		\caption{The temporary logical-AND gate and its Clifford+T gate implementation.  This Clifford+T gate implementation of the temporary logical-AND gate has a T-count of $4$.  $\ket{A}$ is an ancillae in the state $\frac{1}{\sqrt{2}}(\ket{0} + e^{\frac{i \cdot \pi}{4}}\ket{1})$.  Source: \cite{Gidney20184TgateToffolibuild}}
		\label{bi-logical-AND}
	\end{subfigure} \qquad \begin{subfigure}[hb]{3in}
		\small
		\[
		\Qcircuit @C=0.4em @R=0.5em @!R{
			\lstick{\ket{x}} & & \qw &\qw & \qw & \ctrl{1} &\qw & \rstick{\ket{x}} & & & & & & & & & & & & & & & \lstick{\ket{x}} & & \ctrl{1} & \qw & \rstick{\ket{x}}\\
			\lstick{\ket{y}} & & \qw & \qw & \qw & \gate{Z} &\qw & \rstick{\ket{y}} & & & & & & & & & & & & & & & \lstick{\ket{y}} & & \ctrl{1} & \qw & \rstick{\ket{y}} \\
			\lstick{\ket{x \cdot y}} & & \qw &  \gate{H} & \meter & \control \cw \cwx & & & & & & & & & & & & & & & & & \lstick{\ket{x \cdot y}} &  & \qw &\cw & & \\
		}
		\]
		\caption{The uncomputation gate and its Clifford+T gate implementation.  This Clifford+T gate implementation of the uncomputation gate has a T-count of $0$.  Source \cite{Gidney20184TgateToffolibuild}} 
		\label{bi-uncompute-fault-tolerant}
	\end{subfigure} 
	\caption{ Clifford+T gate implementation of the uncomputation gate and logical-AND gate used in this work.}
	\label{QCLA-logic-gates}
\end{figure}

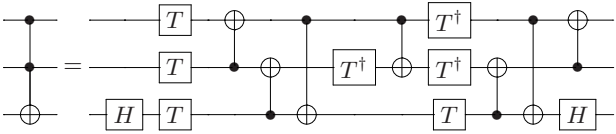
\begin{figure}[tbhp]
	\flushleft
	\small
	\[
	\Qcircuit @C=0.7em @R=0.5em @!R{
		&	&\ctrl{2}		&\qw	&		&		&\qw &\gate{T}	&\qw		&\targ		&\qw			&\ctrl{2}		&\qw		&\ctrl{1}		&\gate{T^{\dag}}	&\qw			&\ctrl{2}		&\targ		&\qw		&\\
		&	&\ctrl{1}		&\qw		& = 	&		&\qw		&\gate{T}	&\qw		&\ctrl{-1}		&\targ		&\qw			&\gate{T^{\dag}}	&\targ		&\gate{T^{\dag}}	&\targ		&\qw			&\ctrl{-1}		&\qw		&\\
		&	&\targ		&\qw		&		&	&\gate{H}		&\gate{T}	&\qw	&\qw		&\ctrl{-1}		&\targ		&\qw	&\qw		&\gate{T}	&\ctrl{-1}		&\targ		&\gate{H}	&\qw		&
	}
	\]
	\caption{The Clifford+T gate implementation of the Toffoli gate design presented in \cite{Maslov}.  This fault tolerant Clifford + T gate implementation of the Toffoli gate has a T-count of $7$. }
	\label{mult:FIG6}
\end{figure}

The Clifford+T gates can be combined to build logic gates which can in turn be used to implement quantum circuits such as the proposed QCLAs.  The proposed QCLA circuits are based on the NOT gate, CNOT gate, Toffoli gate, temporary logical-AND gate and uncomputation gate.  The NOT gate and CNOT gate are members of the Clifford+T gate set.  The Toffoli gate, temporary logical-AND gate and uncomputation gates must be constructed from Clifford+T gates.  We use the Clifford+T implementations of the Toffoli gate designed in \cite{Maslov} for this work (see figure \ref{mult:FIG6}).  Figure \ref{QCLA-logic-gates} shows the temporary logical-AND gate and the uncomputation gates used in this work.  The Clifford+T implementation and the graphical image are shown for each gate.  The design of the logical-AND gate used in this work is presented in \cite{Gidney20184TgateToffolibuild}.    The T-count for these logical gates is as follows:  The Toffoli gate implementation designed in \cite{Maslov} has a T-count of $7$, the temporary logical-AND gate has a T-count of $4$ and the uncomputation gate has a T-count of $0$.

\subsection{Carry Look Ahead Addition}

Carry look ahead addition (CLA) has caught the attention of researchers because carry lookahead addition can perform addition in $\mathcal{O}(log(n))$ time while alternative adders such as ripple carry addition requires $\mathcal{O}(n)$ time.  Thus a CLA circuit can complete the addition operation with a critical path depth of order $\mathcal{O}(log(n))$ whereas the critical path for ripple carry addition has a depth of order $\mathcal{O}(n)$.  The decrease in computation depth comes at the cost of added circuitry.    

Given two $n$ bit inputs $a$ and $b$, the CLA circuit generates the sum of $a$ and $b$ by executing the CLA algorithm illustrated in Figure \ref{CLA-algorithm}.  The CLA algorithm is an established technique to quickly perform addition \cite{Spaniol1981arithmetic} \cite{Parhami2000arithmetic}.  The CLA algorithm can be divided into two basic steps: (i) implement generate ($g_i$) and propagate bits ($p_i$) for each bit of the inputs $a$ and $b$ and (ii) compute the sum $s$ by computing $c_{i+1} = p_{i} \wedge c_{i} \vee g_{i}$ for $1 \leq i \leq n$.

\begin{figure}[hbtp]
	
	\begin{tabular}{ll}
		\\ \midrule
		\multicolumn{2}{l}{\textbf{Algorithm 1:} Carry lookahead Addition }\\ \toprule
		\multicolumn{2}{l}{ \textbf{Function} CLA($a,b$) }\\ 
		\multicolumn{2}{l}{ Requirements: }\\
		\multicolumn{2}{l}{ \qquad //Takes 2 $n$ bit values $a$ and $b$ as input. }\\
		\multicolumn{2}{l}{ \qquad //Returns the sum as an $n+1$ bit number $s$. }\\
		&  \\
		1 & //create carry generate and carry propagate bits. \\
		2 & \\
		3 & \textbf{For} $i = 0 \text{ to } n-1$ \\
		4 & \qquad  $p_i = a_i \oplus  b_i$ \\
		5 &  \qquad $g_i = a_i \wedge b_i $ \\
		6 & \textbf{End} \\
		7 &  \\
		8 & //synthesize sum. \\
		9 & \\
		10 & $c_0 = g_0$ //no carry-in\\
		11 & $s_0 = p_0$ \\ 
		12 & \textbf{For} $i = 1 \text{ to } n-1$ \\
		13 & \qquad $c_i = p_{i-1} \wedge c_{i-1} \vee g_{i-1}$ \\ 
		14 & \qquad $s_i = c_i \oplus a_i \oplus b_i$ \\ 
		15 & \textbf{End} \\ 
		16 & $s_n = p_{n-1} \wedge c_{n-1} \vee g_{n-1}$ \\ 
		17 & \\
		18 & \textbf{Return} $s$ \\ \bottomrule
	\end{tabular}
	\caption{The carry lookahead addition algorithm implemented by the circuits in this work. }
	\label{CLA-algorithm}
\end{figure}

%----------------------------------------------------------------------------------------------------------
\section{Proposed Designs of Out-of-Place QCLA Circuits} 
\label{OOP-QCLA-design}

We now show our proposed out-of-place QCLA circuits.  The proposed QCLA circuits have lower T-count and qubit cost than existing works.  The QCLA designs save T gates by using the temporary logical-AND gate, the existing uncomputation gate or proposed uncomputation gate where possible.  The out of place FT-QCLA1 (or Out-FT-QCLA1) is optimized for T-count.  We also propose an out of place FT-QCLA2 (or Out-FT-QCLA2) which is optimized for qubit cost.  The design methodologies of the proposed QCLAs are generic and each can be used to implement a QCLA circuit of any size.  

The proposed out-of-place QCLA circuits operate as follows: Given 2 values $a$ and $b$ (each $n$ bits wide) stored in quantum registers $A$ and $B$ as well as $n+1$ ancillae stored in register $X$.  $X_0$ is set to $0$ and the remaining locations are set to $A$ $\left( \text{ where } A = \frac{1}{\sqrt{2}}\left(\ket{0} + e^{\frac{i \pi}{4}}\ket{1}\right)\right)$.  Lastly, Out-FT-QCLA2 requires a register $Z$ with $n-w(n)-\lfloor log(n) \rfloor$ elements (where $w(n) = n - \sum_{y=1}^{\infty} \lfloor \frac{n}{2^y} \rfloor$ and is the number of ones in the binary expansion of $n$) all set to $A$ $\left( \text{ where } A = \frac{1}{\sqrt{2}}\left(\ket{0} + e^{\frac{i \pi}{4}}\ket{1}\right)\right)$.  Out-FT-QCLA1 requires a register $Z$ with $3 \cdot n - 2 \cdot w(n) - 2 \lfloor log(n) \rfloor$ ancillae set to $A$.  At the end of computation, $A$ and $B$ are restored to their initial values and $X$ will contain the sum of the addition of $a$ and $b$.  Lastly, $Z$ is transformed into a register of classical states.  These qubits can be restored to computational basis values for reuse as ancillae.  

This Section is organized as follows: The proposed design of Out-FT-QCLA1 is shown in Section \ref{OOP-fault-tolerant-T-gate-QCLA}.  The proposed design of Out-FT-QCLA2 is shown in Section \ref{OOP-fault-tolerant-qubit-QCLA}.      

\subsection{Proposed Out-of-Place FT-QCLA1 Design (Out-FT-QCLA1)}
\label{OOP-fault-tolerant-T-gate-QCLA}

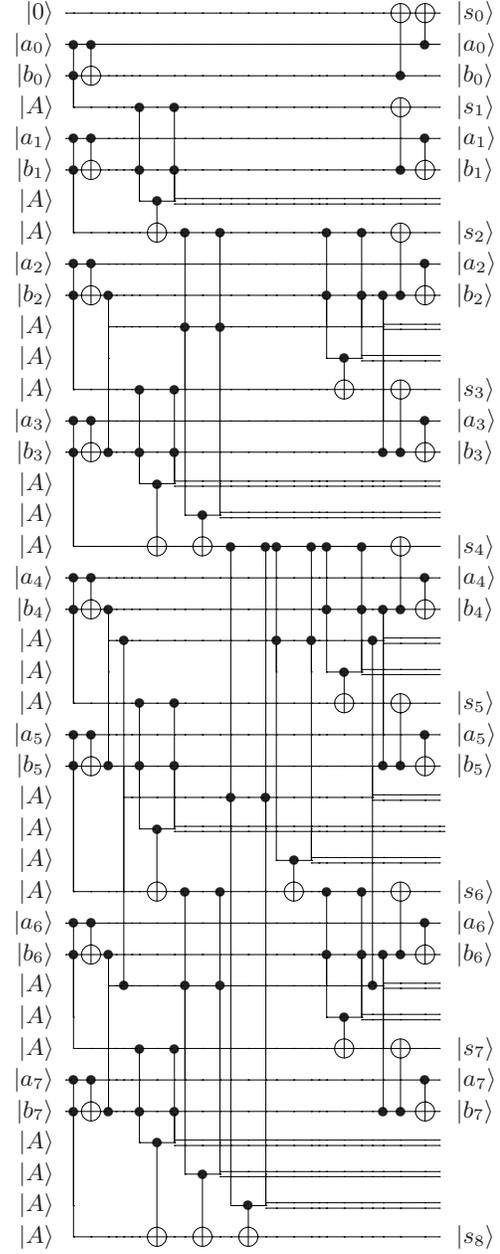
\begin{figure}[htbp]
	\centering
	\small
	\[
	\Qcircuit @C=0.2em @R=0.5em @!R{
		\lstick{\ket{0}} & \qw & \qw & \qw & \qw & \qw &\qw & \qw & \qw & \qw & \qw & \qw & \qw & \qw & \qw & \qw & \qw & \qw & \qw & \qw & \qw & \qw & \qw & \qw & \qw & \targ &\targ & \qw &\rstick{\ket{s_0}} \\
		\lstick{\ket{a_0}} & \ctrl{2} & \ctrl{1} & \qw & \qw & \qw &\qw & \qw & \qw & \qw & \qw & \qw & \qw & \qw & \qw & \qw & \qw & \qw & \qw & \qw & \qw & \qw & \qw & \qw & \qw & \qw & \ctrl{-1}& \qw &\rstick{\ket{a_0}} \\
		\lstick{\ket{b_0}} & \ctrl{1} & \targ & \qw & \qw & \qw &\qw & \qw & \qw & \qw & \qw & \qw & \qw & \qw & \qw & \qw & \qw & \qw & \qw & \qw & \qw & \qw  & \qw & \qw &\qw & \ctrl{-2} & \qw & \qw &\rstick{\ket{b_0}} \\
		\lstick{\ket{A}}   &  & \qw & \qw &\qw & \qw & \qw & \ctrl{3} & \qw &\ctrl{3} & \qw & \qw & \qw & \qw & \qw & \qw & \qw & \qw & \qw & \qw & \qw & \qw & \qw & \qw & \qw & \targ & \qw & \qw  &\rstick{\ket{s_1}} \\
		\lstick{\ket{a_1}} & \ctrl{3} & \ctrl{1} & \qw & \qw & \qw &\qw &  \qw & \qw & \qw & \qw & \qw & \qw & \qw & \qw & \qw & \qw & \qw & \qw& \qw & \qw & \qw & \qw  & \qw & \qw  & \qw&\ctrl{1}& \qw &\rstick{\ket{a_1}} \\ 
		\lstick{\ket{b_1}} & \ctrl{2} & \targ & \qw & \qw & \qw & \qw & \ctrl{1} & \qw & \ctrl{1} & \qw & \qw & \qw & \qw & \qw & \qw & \qw & \qw & \qw & \qw & \qw & \qw & \qw & \qw & \qw &\ctrl{-2} &\targ &\qw &\rstick{\ket{b_1}} \\ 
		\lstick{\ket{A}}   &  &   &  &  &  &  &  &\ctrl{1} &\qw \qwx &\cw  &\cw  &\cw  &\cw &\cw  &\cw  &\cw  &\cw  &\cw  &\cw  &\cw  &\cw  &\cw  &\cw  &\cw  &\cw  &\cw  &\cw       \\
		\lstick{\ket{A}}   & & \qw & \qw & \qw & \qw & \qw & \qw & \targ &\qw  & \ctrl{5} & \qw & \ctrl{5} & \qw & \qw & \qw & \qw & \qw & \qw & \qw &\ctrl{3} &  \qw &\ctrl{3} & \qw &  \qw & \targ & \qw & \qw &\rstick{\ket{s_2}} \\  
		\lstick{\ket{a_2}} & \ctrl{4} & \ctrl{1} & \qw & \qw & \qw & \qw & \qw & \qw & \qw & \qw & \qw & \qw & \qw & \qw & \qw & \qw & \qw & \qw &  \qw & \qw &  \qw & \qw & \qw& \qw  & \qw &\ctrl{1} & \qw &\rstick{\ket{a_2}} \\
		\lstick{\ket{b_2}} & \ctrl{3} & \targ & \ctrl{2} & \qw &\qw & \qw & \qw & \qw & \qw &\qw  &\qw & \qw & \qw & \qw & \qw & \qw & \qw & \qw & \qw & \ctrl{2} & \qw &\ctrl{1} & \qw &\ctrl{4} &\ctrl{-2} &\targ &\qw &\rstick{\ket{b_2}}  \\
		\lstick{\ket{A}}   &  &  &  & \qw &\qw & \qw & \qw & \qw & \qw & \ctrl{6} &\qw &\ctrl{6}  &\qw  & \qw & \qw & \qw & \qw & \qw & \qw &\qw & \qw & \qw & \qw & \qw & \cw & \cw & \cw \\
		\lstick{\ket{A}}   &  &   &  &  &  &  &  &  &  &  &  &  &  &  &  &  &  &  &  &  &\ctrl{1} &\qw \qwx &\cw  &\cw  &\cw &\cw  &\cw   \\
		\lstick{\ket{A}}   &  & \qw & \qw & \qw &\qw &\qw & \ctrl{3} &\qw  & \ctrl{3} &\qw & \qw & \qw & \qw & \qw & \qw & \qw & \qw & \qw & \qw & \qw & \targ & \qw & \qw & \qw & \targ & \qw &\qw &\rstick{\ket{s_3}} \\
		\lstick{\ket{a_3}} & \ctrl{4} & \ctrl{1} & \qw & \qw & \qw &\qw & \qw & \qw & \qw & \qw & \qw &\qw & \qw & \qw & \qw & \qw & \qw & \qw & \qw & \qw & \qw & \qw & \qw & \qw & \qw &\ctrl{1} & \qw &\rstick{\ket{a_3}} \\
		\lstick{\ket{b_3}} & \ctrl{3} & \targ & \ctrl{-5} & \qw &\qw & \qw & \ctrl{1} & \qw & \ctrl{1} & \qw & \qw & \qw & \qw & \qw & \qw & \qw & \qw & \qw & \qw & \qw & \qw &\qw &\qw  &\ctrl{-5}  & \ctrl{-2} &\targ & \qw &\rstick{\ket{b_3}} \\ 
		\lstick{\ket{A}}   &  & & & & & & & \ctrl{2} & \qw \qwx &\cw  &\cw  &\cw  &\cw &\cw  &\cw  &\cw  &\cw &\cw  &\cw  &\cw  &\cw &\cw  &\cw  &\cw  &\cw &\cw &\cw   & \\
		\lstick{\ket{A}}   &  & & & & & & &  &  &  &\ctrl{1}  &\qw \qwx &\cw  &\cw  &\cw  &\cw &\cw  &\cw  &\cw  &\cw &\cw  &\cw  &\cw  &\cw &\cw  &\cw  &\cw     \\
		\lstick{\ket{A}}   &  & \qw & \qw & \qw & \qw & \qw &\qw & \targ &\qw  &\qw  & \targ & \qw & \ctrl{9} & \qw & \ctrl{9} & \ctrl{4} &\qw &\ctrl{4} & \qw & \ctrl{2} &\qw &\ctrl{2} & \qw & \qw & \targ &\qw &\qw &\rstick{\ket{s_4}} \\
		\lstick{\ket{a_4}} & \ctrl{4} & \ctrl{1} & \qw &\qw & \qw & \qw & \qw & \qw & \qw & \qw & \qw & \qw & \qw &\qw & \qw & \qw & \qw & \qw & \qw & \qw & \qw & \qw & \qw & \qw  & \qw &\ctrl{1} &\qw &\rstick{\ket{a_4}} \\
		\lstick{\ket{b_4}} & \ctrl{3} & \targ & \ctrl{2} & \qw & \qw & \qw & \qw & \qw & \qw &\qw & \qw &\qw & \qw &  \qw &\qw & \qw & \qw & \qw & \qw & \ctrl{2} &\qw &\ctrl{2} & \qw & \ctrl{4} & \ctrl{-2} &\targ & \qw &\rstick{\ket{b_4}} \\
		\lstick{\ket{A}}   &  &  &  & \qw & \ctrl{8} & \qw & \qw & \qw &  \qw & \qw & \qw & \qw & \qw &\qw & \qw & \ctrl{7} &\qw & \ctrl{7} & \qw  &  \qw & \qw & \qw & \ctrl{5} & \qw & \cw & \cw & \cw  \\
		\lstick{\ket{A}}   &  &  &  &  &  & & & &  & & & &  & & & &  & & & & \ctrl{1} & \qw \qwx & \cw & \cw & \cw & \cw & \cw &  \\
		\lstick{\ket{A}}   &  & \qw & \qw & \qw & \qw & \qw & \ctrl{3} & \qw &  \ctrl{3} &\qw & \qw & \qw & \qw & \qw &\qw & \qw &\qw & \qw & \qw & \qw &\targ  & \qw &\qw & \qw & \targ &\qw &\qw &\rstick{\ket{s_5}} \\
		\lstick{\ket{a_5}} & \ctrl{5} & \ctrl{1} & \qw & \qw & \qw & \qw & \qw &  \qw &  \qw & \qw & \qw & \qw &\qw & \qw & \qw & \qw &\qw & \qw & \qw &  \qw & \qw & \qw & \qw & \qw & \qw &\ctrl{1} &\qw &\rstick{\ket{a_5}} \\
		\lstick{\ket{b_5}} & \ctrl{4} & \targ & \ctrl{-5} & \qw & \qw & \qw & \ctrl{2} &  \qw & \ctrl{2} & \qw & \qw & \qw &\qw & \qw & \qw & \qw & \qw & \qw &\qw &  \qw & \qw & \qw & \qw & \ctrl{-5} & \ctrl{-2} &\targ &\qw &\rstick{\ket{b_5}} \\
		\lstick{\ket{A}}   &  &  &  &  &  & \qw & \qw & \qw & \qw & \qw & \qw & \qw & \ctrl{13} & \qw & \ctrl{13} & \qw & \qw & \qw &\qw & \qw & \qw &  \qw & \qw & \cw & \cw & \cw & \cw  \\
		\lstick{\ket{A}}   &  &  &  &  &  & & & \ctrl{2} & \qw \qwx & \cw & \cw & \cw & \cw  & \cw & \cw & \cw & \cw  & \cw & \cw & \cw & \cw  & \cw & \cw & \cw & \cw  & \cw & \cw & \cw \\
		\lstick{\ket{A}}   &  &  &  &  &  & & &  & & &  &  &  &  &  & &  \ctrl{1} & \qw \qwx & \cw & \cw & \cw & \cw & \cw & \cw &\cw  &\cw &\cw   \\
		\lstick{\ket{A}}   &  & \qw & \qw & \qw & \qw &\qw & \qw & \targ& \qw & \ctrl{6} & \qw & \ctrl{6} & \qw & \qw & \qw &\qw &\targ & \qw & \qw &  \ctrl{3} &\qw & \ctrl{3} &\qw & \qw & \targ & \qw & \qw &\rstick{\ket{s_6}} \\
		\lstick{\ket{a_6}} & \ctrl{3} & \ctrl{1} & \qw & \qw & \qw & \qw & \qw &  \qw & \qw & \qw & \qw & \qw & \qw & \qw  &\qw & \qw & \qw &  \qw & \qw & \qw & \qw & \qw & \qw & \qw & \qw&\ctrl{1} & \qw &\rstick{\ket{a_6}} \\
		\lstick{\ket{b_6}} & \ctrl{3} & \targ & \ctrl{5} & \qw & \qw & \qw & \qw &  \qw & \qw & \qw &\qw & \qw &  \qw &\qw & \qw &\qw & \qw & \qw &  \qw  &\ctrl{2} & \qw & \ctrl{2} & \qw & \ctrl{3} & \ctrl{-2} &\targ & \qw &\rstick{\ket{b_6}} \\
		\lstick{\ket{A}}   &  &  &  & \qw & \ctrl{-7} & \qw & \qw  &\qw &\qw & \ctrl{6} &  \qw & \ctrl{6} &\qw & \qw &\qw & \qw &\qw & \qw & \qw &  \qw & \qw &\qw & \ctrl{-7} &\qw  & \cw & \cw & \cw   \\
		\lstick{\ket{A}}   &  &   &  &  &  &  &  &  &  &  &  &  &  &  &  &  &  &  &  &  &\ctrl{1} &\qw \qwx & \cw & \cw &\cw  &\cw &\cw  \\
		\lstick{\ket{A}}   &  & \qw & \qw & \qw & \qw & \qw & \ctrl{2}  & \qw &  \ctrl{3} & \qw & \qw & \qw & \qw &\qw & \qw & \qw & \qw & \qw &  \qw & \qw & \targ & \qw & \qw & \qw &\targ & \qw & \qw &\rstick{\ket{s_7}}  \\
		\lstick{\ket{a_7}} & \ctrl{4} & \ctrl{1} & \qw & \qw & \qw & \qw & \qw & \qw &  \qw & \qw & \qw & \qw & \qw & \qw &\qw & \qw & \qw &\qw & \qw & \qw & \qw & \qw & \qw & \qw  & \qw&\ctrl{1} & \qw &\rstick{\ket{a_7}}\\
		\lstick{\ket{b_7}} & \ctrl{4} & \targ & \ctrl{-5} & \qw & \qw & \qw & \ctrl{1} & \qw &  \ctrl{1} & \qw & \qw &\qw & \qw & \qw &\qw & \qw & \qw & \qw & \qw & \qw & \qw & \qw & \qw & \ctrl{-5} & \ctrl{-2} &\targ & \qw &\rstick{\ket{b_7}} \\
		\lstick{\ket{A}}   & &  &  &  &  & & & \ctrl{3} & \qw \qwx & \cw & \cw & \cw & \cw & \cw & \cw & \cw & \cw & \cw & \cw & \cw & \cw & \cw & \cw & \cw & \cw & \cw  &\cw \\
		\lstick{\ket{A}}   & &  &  &  &  & & &  &  & &\ctrl{2} & \qw \qwx & \cw & \cw & \cw & \cw & \cw & \cw & \cw & \cw & \cw & \cw & \cw & \cw & \cw & \cw &\cw \\
		\lstick{\ket{A}}   & &  &  &  &  & & &  &  & & &  &  & \ctrl{1} & \qw \qwx & \cw & \cw & \cw & \cw & \cw & \cw & \cw & \cw & \cw & \cw & \cw & \cw \\
		\lstick{\ket{A}}   &  & \qw & \qw &\qw & \qw & \qw &\qw &\targ &\qw & \qw & \targ & \qw & \qw & \targ & \qw & \qw & \qw & \qw & \qw & \qw &\qw & \qw &\qw & \qw &\qw & \qw &\qw &\rstick{\ket{s_8}}  \\
	}
	\]
	\caption{Proposed out-of-place FT-QCLA1 (Out-FT-QCLA1) circuit for the case of adding two $8$ bit values $a$ and $b$.}
	\label{OOP-TgateOptimized}
\end{figure}

The out-of-place FT-QCLA1 (Out-FT-QCLA1) is optimized for T-count.  The proposed QCLA is based on the quantum NOT gate, CNOT gate, logical AND gate and uncomputation gate.  The steps of the proposed design methodology to realize Out-FT-QCLA1 are shown along with an illustrative example of the QCLA circuit in Figure \ref{OOP-TgateOptimized}.

\begin{itemize}
	
	\item \textbf{Step 1:} For \textit{i} = 0 to n-1, apply the logical AND gate at locations A[\textit{i}], B[\textit{i}] and an ancillae such that locations A[\textit{i}] and B[\textit{i}] are unchanged.  The ancillae will have the result of computation.  The ancillae will be renamed to the value g[\textit{i}, \textit{i} +1]. 
	
	\item \textbf{Step 2:} For \textit{i} = 1 to n-1. At the locations A[\textit{i}] and B[\textit{i}] apply a CNOT gate such that A[\textit{i}] is unchanged and location B[\textit{i}] will have the result of computation. Location B[\textit{i}] will be renamed p[\textit{i}, \textit{i}+1]. \\
	
	\item \textbf{Step 3 (P-rounds):} We use a nested loop in this step. For $t = 1 \text{ to } \lfloor log(n) \rfloor-1$ and For $m  = 1 \text{ to } \lfloor \frac{n}{2^t} \rfloor - 1$:
	
	At locations p[\textit{j},\textit{l}], p[\textit{l}, \textit{k}] and at an ancillae apply a logical AND gate such that locations p[\textit{j},\textit{l}] and p[\textit{l}, \textit{k}] are unchanged and the ancillae will have the result of computation.  The ancillae will be renamed p[\textit{j},\textit{k}].  The equation for indexes are \textit{j} = $2^t \cdot m$, \textit{k} = $2^t \cdot m + 2^t$, and \textit{l} = $2^t \cdot m + 2^{t-1}$, respectively.  
	
	\item \textbf{Step 4 (G-rounds):} We use a nested loop in this step. For $t = 1 \text{ to } \lfloor log(n) \rfloor$ and For $m  = 0 \text{ to } \lfloor \frac{n}{2^t} \rfloor - 1$:
	
	At locations g[\textit{j}, \textit{l}], p[\textit{l}, \textit{k}], and g[\textit{l}, \textit{k}] apply a Logical AND gate and uncomputation gate pair such that locations g[\textit{j}, \textit{l}] and p[\textit{l}, \textit{k}] are unchanged and location g[\textit{l}, \textit{k}] will have the result of computation.  Location g[\textit{l}, \textit{k}] is renamed to g[\textit{j}, \textit{k}].  The equation for indexes are \textit{j} = $2^t \cdot m$, \textit{k} = $2^t \cdot m + 2^t$, and \textit{l} = $2^t \cdot m + 2^{t-1}$, respectively.

	\item \textbf{Step 5 (C-rounds):} We use a nested loop in this step. For $t = \lfloor log \left(\frac{2\cdot n}{3} \right) \rfloor \text{ to } 1$ and For $m  = 1 \text{ to } \lfloor \frac{(n - 2^{t-1})}{2^t} \rfloor$:
	
	At locations g[0, \textit{l}], p[\textit{l},\textit{k}], and g[\textit{l},\textit{k}] apply a Logical AND gate and uncomputation gate pair such that locations g[0, \textit{l}] and p[\textit{l},\textit{k}] are unchanged and location g[\textit{l},\textit{k}] will have the result of computation.  The equation for indexes are \textit{l} = $2^t \cdot m$ and \textit{k} = $2^t \cdot m + 2^{t-1}$.  
	
	%%%%%%%%%%%%%%%%%%%%%%%%%%%%%%%%%%%%%%%%%%%
	
	\item \textbf{Step 6 (P-erase-rounds):}  We use a nested loop in this step. For $t = \lfloor log(n) \rfloor - 1 \text{ to }1$ and For $m  = 1 \text{ to } \lfloor \frac{n}{2^t} \rfloor - 1$:
	
	At locations  p[\textit{j}, \textit{l}], p[\textit{l}, \textit{k}], and p[\textit{j}, \textit{k}] apply a uncomputation gate such that locations p[\textit{j}, \textit{l}]] and p[\textit{l},\textit{k}] are unchanged and location p[\textit{j}, \textit{k}] will be restored to its original value.  The equation for indexes \textit{j} = $2^t \cdot m$, \textit{k} = $2^t \cdot m + 2^t$, and \textit{l} = $2^t \cdot m + 2^{t-1}$.
	
	\item \textbf{Step 7:} This step has two sub-steps:
	
	\begin{itemize}
		
		\item \textbf{Sub-step 1:} For \textit{i}  = 1 to n-1, At locations p[\textit{i}, \textit{i} +1] and g[0, \textit{i}] apply a CNOT gate such that location p[\textit{i}, \textit{i} +1] is unchanged and g[0, \textit{i}] has the result of computation.  After this step, location g[0, \textit{i}] will have the sum bit $s_{{\textit{i}}}$.  The sum bit $s_{\textit{n}}$ is at location  g[\textit{0}, \textit{n}].  
		
		\item \textbf{Sub-step 2:} At locations p[\textit{0}, \textit{1}] and an ancillae apply a CNOT gate such that location p[\textit{0}, \textit{1}] and the ancillae will have the value B[\textit{0}].  %FIX!!!!!!!!!!!!!!!!!!!!!!!!!!!!!!!!!!!!!!!!!!!!
		
	\end{itemize}
	
	\item \textbf{Step 8:} This step has two sub-steps:
	
	\begin{itemize}
		
		\item \textbf{Sub-step 1:} For \textit{i}  = 1 to n-1, At locations  p[\textit{i}, \textit{i} +1] and A[\textit{i}] apply a CNOT gate such that A[\textit{i}] is unchanged and p[\textit{i}, \textit{i} +1] is restored to its original value ($b_{\textit{i}}$).
		
		\item \textbf{Sub-step 2:}  At locations  Z and A[\textit{0}] apply a CNOT gate such that A[\textit{0}] is unchanged and Z will have the sum bit $s_{{\textit{0}}}$
		
	\end{itemize}
	
\end{itemize}

\subsection{Proposed Out-of-Place FT-QCLA2 Design (Out-FT-QCLA2)}
\label{OOP-fault-tolerant-qubit-QCLA}

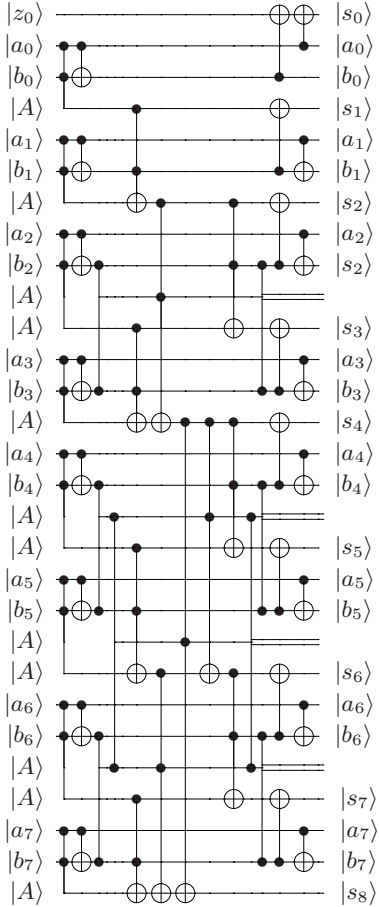
\begin{figure}[htbp]
	\centering
	\small
	\[
	\Qcircuit @C=0.2em @R=0.5em @!R{
		\lstick{\ket{z_0}} & \qw & \qw & \qw & \qw & \qw & \qw & \qw & \qw & \qw & \qw & \qw & \qw & \qw & \targ & \targ & \qw & \rstick{\ket{s_0}}\\
		\lstick{\ket{a_0}} & \ctrl{2} & \ctrl{1} & \qw & \qw & \qw & \qw & \qw & \qw & \qw & \qw & \qw & \qw & \qw & \qw & \ctrl{-1} & \qw & \rstick{\ket{a_0}}\\
		\lstick{\ket{b_0}} & \ctrl{1} & \targ & \qw & \qw & \qw & \qw & \qw & \qw & \qw & \qw & \qw & \qw & \qw & \ctrl{-2} & \qw & \qw & \rstick{\ket{b_0}} \\
		\lstick{\ket{A}}   &  & \qw & \qw &\qw & \qw & \qw & \ctrl{3} & \qw & \qw & \qw & \qw & \qw & \qw & \targ & \qw & \qw & \rstick{\ket{s_1}}\\
		\lstick{\ket{a_1}} & \ctrl{2} & \ctrl{1} & \qw & \qw & \qw & \qw & \qw & \qw & \qw & \qw & \qw & \qw & \qw & \qw & \ctrl{1} & \qw & \rstick{\ket{a_1}}\\ 
		\lstick{\ket{b_1}} & \ctrl{1} & \targ & \qw & \qw & \qw & \qw & \ctrl{1} & \qw & \qw & \qw & \qw & \qw &\qw & \ctrl{-2} & \targ & \qw & \rstick{\ket{b_1}}\\ 
		\lstick{\ket{A}}   & & \qw & \qw & \qw & \qw & \qw & \targ & \ctrl{4} & \qw & \qw & \ctrl{3} & \qw & \qw & \targ & \qw & \qw & \rstick{\ket{s_2}}\\  
		\lstick{\ket{a_2}} & \ctrl{2} & \ctrl{1} & \qw & \qw & \qw & \qw & \qw & \qw & \qw & \qw & \qw & \qw & \qw & \qw & \ctrl{1} & \qw & \rstick{\ket{a_2}}\\
		\lstick{\ket{b_2}} & \ctrl{2} & \targ & \ctrl{2} & \qw &\qw & \qw & \qw & \qw & \qw & \qw & \ctrl{2} & \qw & \ctrl{2} & \ctrl{-2} & \targ & \qw & \rstick{\ket{s_2}}\\
		\lstick{\ket{A}}   &  &  &  & \qw &\qw & \qw & \qw & \ctrl{4} & \qw & \qw & \qw &\qw & \qw &\cw  &\cw &\cw &\cw & \\
		\lstick{\ket{A}}   &  & \qw & \qw & \qw &\qw & \qw & \ctrl{3} & \qw & \qw & \qw & \targ &\qw & \qw & \targ & \qw & \qw & \rstick{\ket{s_3}}\\
		\lstick{\ket{a_3}} & \ctrl{2} & \ctrl{1} & \qw & \qw &\qw & \qw & \qw & \qw & \qw & \qw & \qw &\qw & \qw & \qw & \ctrl{1} & \qw & \rstick{\ket{a_3}}\\
		\lstick{\ket{b_3}} & \ctrl{1} & \targ & \ctrl{-2} & \qw &\qw & \qw & \ctrl{1} & \qw & \qw & \qw & \qw & \qw & \ctrl{-2} & \ctrl{-2} &\targ & \qw & \rstick{\ket{b_3}}\\ 
		\lstick{\ket{A}}   &  & \qw & \qw & \qw & \qw & \qw & \targ & \targ & \ctrl{7} & \ctrl{4} &\ctrl{3} & \qw & \qw & \targ & \qw & \qw & \rstick{\ket{s_4}}\\
		\lstick{\ket{a_4}} & \ctrl{2} & \ctrl{1} & \qw &\qw & \qw & \qw & \qw & \qw & \qw & \qw & \qw & \qw & \qw & \qw & \ctrl{1} & \qw & \rstick{\ket{a_4}}\\
		\lstick{\ket{b_4}} & \ctrl{2} & \targ & \ctrl{2} & \qw & \qw & \qw & \qw & \qw & \qw &\qw & \ctrl{2} & \qw & \ctrl{2} & \ctrl{-2} & \targ & \qw & \rstick{\ket{b_4}}\\
		\lstick{\ket{A}}   &  &  &  & \qw & \ctrl{4} & \qw & \qw & \qw &  \qw & \ctrl{5} & \qw &\ctrl{4} &\qw &\cw &\cw &\cw &\cw &\\
		\lstick{\ket{A}}   &  & \qw & \qw & \qw & \qw & \qw & \ctrl{3} & \qw &  \qw &\qw & \targ & \qw & \qw & \targ & \qw & \qw & \rstick{\ket{s_5}}\\
		\lstick{\ket{a_5}} & \ctrl{2} & \ctrl{1} & \qw & \qw & \qw & \qw & \qw &  \qw &  \qw & \qw & \qw & \qw & \qw & \qw & \ctrl{1} & \qw &\rstick{\ket{a_5}} \\
		\lstick{\ket{b_5}} & \ctrl{2} & \targ & \ctrl{-2} & \qw & \qw & \qw & \ctrl{2} &  \qw & \qw & \qw & \qw & \qw & \ctrl{-2} & \ctrl{-2} & \targ & \qw & \rstick{\ket{b_5}}\\
		\lstick{\ket{A}}   &  &  &  &  &  & \qw & \qw & \qw &  \ctrl{8} & \qw & \qw & \qw &\cw &\cw &\cw  &\cw &\\
		\lstick{\ket{A}}   &  & \qw & \qw & \qw & \qw & \qw & \targ& \ctrl{3} & \qw & \targ & \ctrl{3} & \qw & \qw & \targ & \qw & \qw & \rstick{\ket{s_6}}\\
		\lstick{\ket{a_6}} & \ctrl{2} & \ctrl{1} & \qw & \qw & \qw & \qw & \qw &  \qw & \qw & \qw & \qw & \qw & \qw & \qw & \ctrl{1} & \qw & \rstick{\ket{a_6}}\\
		\lstick{\ket{b_6}} & \ctrl{2} & \targ & \ctrl{2} & \qw & \qw & \qw & \qw &  \qw & \qw & \qw & \ctrl{2} & \qw &\ctrl{2} & \ctrl{-2} & \targ & \qw & \rstick{\ket{b_6}}\\
		\lstick{\ket{A}}   &  &  &  & \qw & \ctrl{-4} & \qw & \qw  & \ctrl{4} &  \qw & \qw & \qw &\ctrl{-4} & \qw &\cw &\cw &\cw &\cw &\\
		\lstick{\ket{A}}   &  & \qw & \qw & \qw & \qw & \qw & \ctrl{2}  & \qw &  \qw & \qw & \targ & \qw & \qw & \targ & \qw & \qw & \qw & \rstick{\ket{s_7}}\\
		\lstick{\ket{a_7}} & \ctrl{2} & \ctrl{1} & \qw & \qw & \qw & \qw & \qw & \qw &  \qw & \qw & \qw & \qw & \qw & \qw & \ctrl{1} & \qw & \qw & \rstick{\ket{a_7}}\\
		\lstick{\ket{b_7}} & \ctrl{1} & \targ & \ctrl{-2} & \qw & \qw & \qw & \ctrl{1} & \qw &  \qw & \qw & \qw & \qw & \ctrl{-2} & \ctrl{-2} & \targ & \qw & \qw & \rstick{\ket{b_7}}\\
		\lstick{\ket{A}}   &  & \qw & \qw &\qw & \qw & \qw &\targ &\targ &\targ & \qw & \qw & \qw & \qw & \qw & \qw & \qw & \qw &\rstick{\ket{s_8}} \\
	}
	\]
	\caption{Proposed out-of-place FT-QCLA2 (Out-FT-QCLA2) for the case of adding two $8$ bit values $a$ and $b$.}
	\label{OOP-QubitOptimized}
\end{figure}

Out-FT-QCLA2 is optimized for qubit cost.  A trade-off for the reduced qubit count is an increase in the T gate cost of Out-FT-QCLA2.  We reduce the qubit cost by replacing logical AND gate and uncomputation gate pairs with alternative Toffoli gate implementations such as the design in \cite{Maslov} were appropriate.  To retain a low T-count, Out-FT-QCLA2 does still incorporate logical AND gate and uncomputation gate pairs.  We save qubits because the logical AND gate and uncomputation gate pairs have a qubit cost of $4$ while implementations of the Toffoli gate (like the design in \cite{Maslov}) have a qubit cost of $3$.  However, alternative Toffoli gate implementations have higher T-counts (such as the design in \cite{Maslov} with a T-count of $7$).  The design methodology for Out-FT-QCLA2 is identical to the methodology for Out-FT-QCLA1 except Step 4 and Step 5.  In Step 4 and Step 5, Toffoli gates based on the design in \cite{Maslov} are used.  An illustrative example of Out-FT-QCLA2 is in Figure \ref{OOP-QubitOptimized}.

\section{Performance of Proposed out-of-place QCLA circuits}
\label{OOP-QCLA-performance}

\begin{table*}[tbhp]
	\centering
	\caption{Cost Comparison of Out-of-place QCLAs}
	%\begin{tabular}{ |p{8cm}||p{8cm}|p{8cm}|p{8cm}| }
	\begin{tabular}{lccc}
		\\ \midrule												
		Design	&	& T-count Equation	&	Qubit Equation \\		\cmidrule{1-1} \cmidrule{3-4}			
		Draper et al. (\cite{Draper2006qcla})	&	& $35n-21w(n)-21\lfloor log(n)\rfloor-7$ &	$4n-w(n)-\lfloor log(n)\rfloor+1$                       	\\
		Trisetyarso et al. (\cite{Trisetyarso2009qcla})	 &  &   $35n-21w(n)-21\lfloor log(n)\rfloor-7$  &     $4n-w(n)-\lfloor log(n)\rfloor+1$    \\
		Thapliyal et al. (\cite{Thapliyal2013qcla} )		&   &   $35n-14$      &     $4n+1$           \\
		Babu et. al. (\cite{Babu2013QCLAoutofplace})$^1$ & & $54 \cdot n$ & $12 \cdot n + 1$  \\
		Lisa et al. (\cite{Lisa2015QCLAoutofplace})$^1$ & & $26 \cdot n$ &  $6 \cdot n + 1$ \\
		Out-FT-QCLA1$^2$ &   &$16n - 8w(n) -  8\lfloor log(n) \rfloor - 4$ &   $ 6n-2w(n)-2\lfloor log(n) \rfloor $  \\
		Out-FT-QCLA2$^3$ &   & $22n-11w(n)-11\lfloor log(n)\rfloor-7$  &    $4n-w(n)-\lfloor log(n) \rfloor + 1$  \\ \bottomrule	
		\multicolumn{4}{l}{$w(n) = n - \sum_{y=1}^{\infty} \bigl\lfloor \frac{n}{2^y} \bigr\rfloor$} \\
		\multicolumn{4}{p{3in}}{$^1$ Circuits modified to remove garbage output.  We use the methodology in \cite{Bennett1973trashremoval} to remove the garbage output.}\\
		\multicolumn{4}{l}{$^2$ Out-FT-QCLA1 is optimized emphasizing T-count.} \\
		\multicolumn{4}{l}{$^3$ Out-FT-QCLA2 is optimized emphasizing qubit cost.} \\
	\end{tabular}
	\label{Equations for Out-of-place T-count}
\end{table*}

\subsection{T-count analysis of Out-FT-QCLA-1}

The T-count of Out-FT-QCLA-1 is shown for each Step of the proposed design methodology.  Total T-count is determined by summing the T-count for each Step of the proposed design methodology.  This design uses logical AND gate and uncomputation gate pairs to implement the Toffoli gate as shown in \cite{Cody2012toffoligatebuild} and \cite{Gidney20184TgateToffolibuild}.  The pair has a T-count of 4.  The total T-count is $16 \cdot n - 8 \cdot w(n) -  8 \cdot \lfloor log(n) \rfloor - 4$ where $w(n) = n - \sum_{y=1}^{\infty} \bigl\lfloor \frac{n}{2^y} \bigr\rfloor$.

\begin{itemize}
	\item Step 1 uses $n$ logical AND gates.  The T-count for this step is $4 \cdot n$.
	\item Step 2 does not need T gates. 
	\item Step 3 uses $n - w(n) - \lfloor log(n) \rfloor$ logical AND gates.  The T-count for this step is $4 \cdot (n - w(n) - \lfloor log(n) \rfloor)$. 
	\item Step 4 uses $n - w(n)$ logical AND gate and uncomputation gate pairs.  The T-count for this step is $4 \cdot (n - w(n))$. 
	\item Step 5 uses $n - \lfloor log(n) \rfloor - 1$ logical AND gate and uncomputation gate pairs.  The T-count for this step is $4 \cdot (n - \lfloor log(n) \rfloor - 1)$. 
	\item Steps 6 through 8 does not need T gates.
\end{itemize}

\subsection{T-count analysis of Out-FT-QCLA-2}

The T-count of Out-FT-QCLA-2 is shown for each Step of the proposed design methodology.  Total T-count is determined by summing the T-count for each Step of the proposed design methodology.  This design uses the Toffoli gate implementation in \cite{Maslov} which has a T-count of $7$.  The total T-count is $22 \cdot n - 11 \cdot w(n) -  11 \cdot \lfloor log(n) \rfloor - 7$ where $w(n) = n - \sum_{y=1}^{\infty} \bigl\lfloor \frac{n}{2^y} \bigr\rfloor$.

\begin{itemize}
	\item Step 1 uses $n$ logical AND gates.  The T-count for this step is $4 \cdot n$.
	\item Step 2 does not need T gates. 
	\item Step 3 uses $n - w(n) - \lfloor log(n) \rfloor$ logical AND gates.  The T-count for this step is $4 \cdot (n - w(n) - \lfloor log(n) \rfloor)$. 
	\item Step 4 uses $n - w(n)$ Toffoli gates.  The T-count for this step is $7 \cdot (n - w(n))$. 
	\item Step 5 uses $n - \lfloor log(n) \rfloor - 1$ Toffoli gates.  The T-count for this step is $7 \cdot (n - \lfloor log(n) \rfloor - 1)$. 
	\item Steps 6 through 8 does not need T gates.
\end{itemize}

\subsection{Cost Comparison of Proposed Out-of-place QCLAs}

\subsubsection{Cost Comparison in Terms of T-count}		

Table \ref{Equations for Out-of-place T-count} indicates that all proposed out-of-place QCLAs have T-count costs of order $\mathcal{O}(n)$.  The existing works also have T-counts of order $\mathcal{O}(n)$.  All proposed QCLAs have a reduced T-count compared to the existing work.  Of the proposed QCLAS, Out-FT-QCLA1 requires the fewest T gates.   

The Out-FT-QCLA1 requires roughly $70.37 \%$ fewer T gates than the design by Babu et al., $38.46 \%$ fewer T gates than the design by Lisa et al., $54.29 \%$ fewer T gates than the designs by Draper et al., Thapliyal et al. and Trisetyarso et al.  

The proposed Out-FT-QCLA2 requires $59.26 \%$ fewer T gates than the design by Babu et al., $15.38 \%$ fewer T gates than the design by Lisa et al., $37.14 \%$ fewer T gates than the designs by Draper et al., Thapliyal et al. and Trisetyarso et al.  

\subsubsection{Cost Comparison in Terms of Qubits}	

Table \ref{Equations for Out-of-place T-count} indicates that all proposed out-of-place QCLAs have qubit costs of order $\mathcal{O}(n)$.  The existing works also have a T-count of order $\mathcal{O}(n)$.  Out-FT-QCLA1 requires $2 \cdot n + w(n)+\lfloor log(n)\rfloor - 1$ additional ancillae compared to Out-FT-QCLA2.  The added qubit cost of Out-FT-QCLA1 illustrates the trade-off between qubits and T gates that occurs when logical AND gate and uncomputation gate pairs are replaced by Toffoli gates in the QCLA implementation.

Out-FT-QCLA2 requires $w(n)+\lfloor log(n)\rfloor$ fewer qubits than the design by Thapliyal et al. and has the same qubit cost as the designs by Draper et al. and Trisetyarso et al.  Further, the proposed Out-FT-QCLA2 requires $8 \cdot n + w(n)+\lfloor log(n)\rfloor$ fewer qubits than the design by Babu et al. and $2 \cdot n + w(n)+\lfloor log(n)\rfloor$ fewer qubits than the design by Lisa et al.  In contrast, the out-of-place Out-FT-QCLA1 requires $6 \cdot n + 2 \cdot w(n) + 2 \cdot \lfloor log(n) \rfloor + 1$ fewer qubits than the design by Babu et al. and $2 \cdot w(n) + 2 \cdot \lfloor log(n) \rfloor + 1$ fewer qubits than the work by Lisa et al.     

%%%%%%%%%%%%%%%%%%%%%%%%%%%%%%%%%%%%%%%%%%%%%%%%%%%%%%%%%%%%%%%%%%%%%%%%%%%%%%%%%%%%%%%%%%%%%%%%%%%%%%%%%%%%%%%%%%%%%%
%\section{Proposed Near Term In-Place QCLA Circuit and Proposed Fault Tolerant In-Place QCLA Circuit}
\section{Proposed In-Place QCLA Circuits} 
\label{IP-QCLA-design}

We now show our proposed in-place quantum carry lookahead (QCLA) circuits.  The proposed QCLA circuits have lower T-count and qubit costs than existing works.  The proposed in-place FT-QCLA1 (or In-FT-QCLA1) and the proposed in-place FT-QCLA2 (or In-FT-QCLA2) save T gates by using the temporary logical-AND gate and uncomputation gate where possible.  In-FT-QCLA2 is optimized for qubit cost and the In-FT-QCLA1 is optimized for T gates.  The methodologies of the proposed in-place QCLAs are generic and each can be used to implement a QCLA circuit of any size.  

The proposed QCLA circuits operate as follows: Given 2 values $a$ and $b$ (each $n$ bits wide) stored in quantum registers $A$, $B$ and an $n$ bit register $Z$ of ancillae set to $a \text{ where } A = \frac{1}{\sqrt{2}}\left(\ket{0} + e^{\frac{i \pi}{4}}\ket{1}\right)$.  Lastly, In-FT-QCLA2 QCLA requires a $n-w(n)-\lfloor log(n) \rfloor$ (where $w(n) = n - \sum_{y=1}^{\infty} \bigl\lfloor \frac{n}{2^y} \bigr\rfloor$ and is the number of ones in the binary expansion of $n$) bit register of ancillae ($X$) set to $\frac{1}{\sqrt{2}}\left(\ket{0} + e^{\frac{i \pi}{4}}\ket{1}\right)$.  In contrast, register $X$ of the In-FT-QCLA1 has $3 \cdot n - 2 \cdot w(n) - 2 \lfloor log(n) \rfloor$ ancillae set to $A$.  At the end of computation, $A$ is restored to their initial values and $B$ will contain sum bits $0$ through $n-1$ of the addition of $a$ and $b$.  For each QCLA, at the end of computation, $Z$ will contain the sum bit $s_n$ and the remaining locations in $Z$ are transformed into a register of classical states.  These qubits can be restored to initial values for reuse as ancillae.  In-FT-QCLA1 and In-FT-QCLA2 will transform the remaining locations in $X$ to a register of classical states.  These qubits can be restored to initial values for reuse as ancillae.  This Section is organized as follows: In-FT-QCLA1 is shown in Section \ref{IP-fault-tolerant-T-gates-QCLA}.  The In-FT-QCLA2 is shown in Section \ref{IP-fault-tolerant-QCLA}.      
%--------------------------------------------------==================================------------------------------

\subsection{Methodology of the Proposed In-Place FT-QCLA1 (In-FT-QCLA1)}
\label{IP-fault-tolerant-T-gates-QCLA}

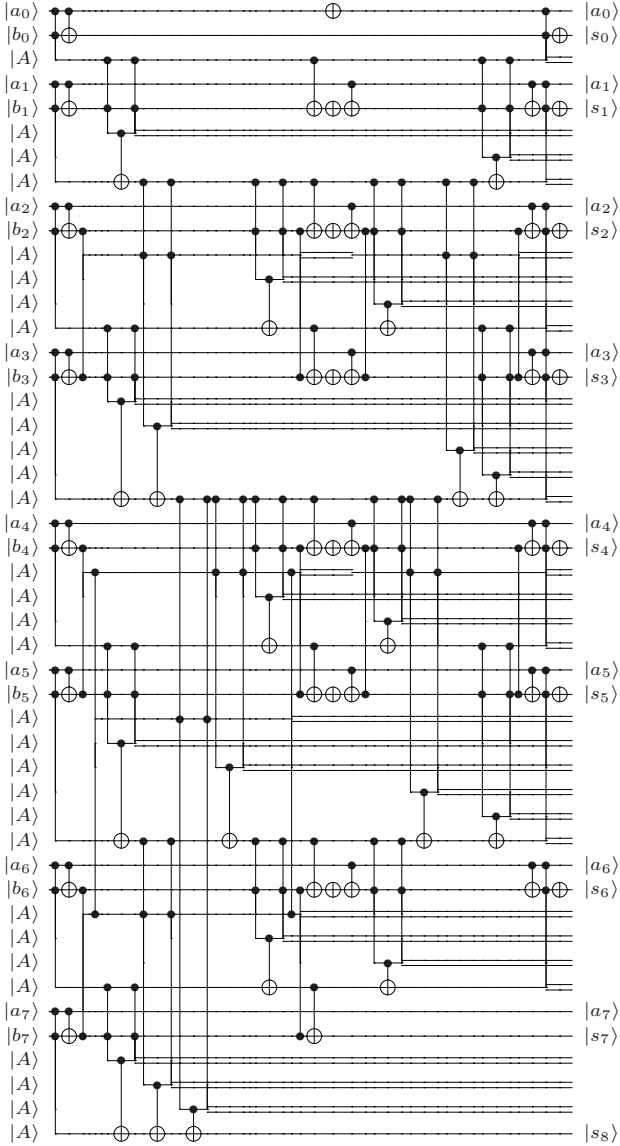
\begin{figure}[hbtp]
	\centering
	\scriptsize
	\[
	\Qcircuit @C=0.2em @R=0.5em @!R{
		\lstick{\ket{a_0}} & \ctrl{2} & \ctrl{1} & \qw & \qw & \qw &\qw & \qw & \qw & \qw & \qw & \qw & \qw & \qw & \qw & \qw & \qw & \qw & \qw & \qw & \qw & \qw & \qw & \qw & \qw &\qw & \targ &\qw & \qw & \qw & \qw & \qw & \qw & \qw & \qw & \qw & \qw & \qw & \qw & \qw & \qw & \qw & \qw &\ctrl{2} & \qw & \qw &\rstick{\ket{a_0}} \\
		\lstick{\ket{b_0}} & \ctrl{1} & \targ & \qw & \qw & \qw &\qw & \qw & \qw & \qw & \qw & \qw & \qw & \qw & \qw & \qw & \qw & \qw & \qw & \qw & \qw & \qw & \qw & \qw &\qw & \qw & \qw & \qw & \qw & \qw & \qw & \qw & \qw & \qw & \qw & \qw & \qw & \qw & \qw & \qw & \qw & \qw & \qw & \ctrl{1} & \targ & \qw &\rstick{\ket{s_0}} \\
		\lstick{\ket{A}}   &  & \qw & \qw &\qw & \qw & \qw & \ctrl{3} & \qw &\ctrl{3} & \qw & \qw & \qw & \qw & \qw & \qw & \qw & \qw & \qw & \qw & \qw & \qw & \qw & \qw & \qw &  \ctrl{2} & \qw & \qw & \qw & \qw &  \qw & \qw & \qw & \qw & \qw & \qw & \qw & \qw & \ctrl{2} & \qw & \ctrl{3} & \qw  &\qw & \qw \qwx &\cw &\cw \\
		\lstick{\ket{a_1}} & \ctrl{3} & \ctrl{1} & \qw & \qw & \qw &\qw &  \qw & \qw & \qw & \qw & \qw & \qw & \qw & \qw & \qw & \qw & \qw & \qw& \qw & \qw & \qw & \qw  & \qw & \qw & \qw & \qw &\ctrl{1} & \qw & \qw & \qw & \qw & \qw & \qw & \qw & \qw & \qw & \qw & \qw & \qw & \qw &\qw &\ctrl{1} &\ctrl{2} &\qw & \qw &\rstick{\ket{a_1}} \\ 
		\lstick{\ket{b_1}} & \ctrl{3} & \targ & \qw & \qw & \qw & \qw & \ctrl{1} & \qw & \ctrl{1} & \qw & \qw & \qw & \qw & \qw & \qw & \qw & \qw & \qw & \qw & \qw & \qw & \qw & \qw & \qw & \targ & \targ & \targ & \qw & \qw & \qw & \qw & \qw & \qw & \qw & \qw  & \qw & \qw & \ctrl{2} & \qw & \ctrl{2} & \qw & \targ &\ctrl{2} &\targ & \qw &\rstick{\ket{s_1}} \\ 
		\lstick{\ket{A}}   &  &   &  &  &  &  &  &\ctrl{2} &\qw \qwx &\cw  &\cw  &\cw  &\cw &\cw  &\cw  &\cw  &\cw  &\cw  &\cw  &\cw  &\cw  &\cw  &\cw  &\cw  &\cw  &\cw  &\cw  &\cw  &\cw  &\cw  &\cw  &\cw  &\cw  &\cw  &\cw  &\cw  &\cw &\cw  &\cw  &\cw  &\cw  &\cw &\cw  &\cw  &\cw  \\
		\lstick{\ket{A}}   &  &   &  &  &  &  &  & & &  &  &  &  &  &  &  &  &  &  &  &  &    &  &  &  &  &  &  &  &  &  &  &  &  &  &  & & &\ctrl{1} &\qw \qwx  &\cw  &\cw  &\cw  &\cw &\cw  \\
		\lstick{\ket{A}}   & & \qw & \qw & \qw & \qw & \qw & \qw & \targ &\qw  & \ctrl{5} & \qw & \ctrl{5} & \qw & \qw & \qw & \qw & \qw & \qw & \qw &\ctrl{3} &  \qw &\ctrl{3} & \qw &  \qw &  \ctrl{2} & \qw & \qw & \qw   &\ctrl{4} &  \qw &\ctrl{4} & \qw & \qw & \qw &\ctrl{5} &  \qw &\ctrl{5} & \qw & \targ &\qw & \qw & \qw  & \qw \qwx &\cw &\cw \\  
		\lstick{\ket{a_2}} & \ctrl{4} & \ctrl{1} & \qw & \qw & \qw & \qw & \qw & \qw & \qw & \qw & \qw & \qw & \qw & \qw & \qw & \qw & \qw & \qw &  \qw & \qw &  \qw & \qw & \qw & \qw & \qw &\qw &\ctrl{1} & \qw &  \qw & \qw & \qw & \qw & \qw & \qw & \qw & \qw & \qw & \qw & \qw & \qw & \qw &\ctrl{1} &\ctrl{3} & \qw &\qw &\rstick{\ket{a_2}} \\
		\lstick{\ket{b_2}} & \ctrl{4} & \targ & \ctrl{2} & \qw &\qw & \qw & \qw & \qw & \qw &\qw  &\qw & \qw & \qw & \qw & \qw & \qw & \qw & \qw & \qw & \ctrl{2} & \qw &\ctrl{1} & \qw &\ctrl{4} & \targ & \targ & \targ & \ctrl{4} & \ctrl{3} & \qw &\ctrl{3} & \qw & \qw & \qw & \qw & \qw & \qw & \qw & \qw & \qw &\ctrl{4} &\targ &\ctrl{5} & \targ & \qw &\rstick{\ket{s_2}} \\
		\lstick{\ket{A}}   &  &  &  & \qw &\qw & \qw & \qw & \qw & \qw & \ctrl{7} &\qw &\ctrl{7}  &\qw  & \qw & \qw & \qw & \qw & \qw & \qw &\qw & \qw & \qw & \qw & \qw & \cw & \cw & \cw & \qw & \qw & \qw & \qw & \qw & \qw & \qw &\ctrl{8} &  \qw &\ctrl{8} & \qw & \qw & \qw & \qw & \cw & \cw & \cw & \cw &\\
		\lstick{\ket{A}}   &  &   &  &  &  &  &  &  &  &  &  &  &  &  &  &  &  &  &  &  &\ctrl{2} &\qw \qwx &\cw  &\cw  &\cw  &\cw  &\cw  &\cw  &\cw  &\cw &\cw  &\cw  &\cw  &\cw &\cw  &\cw  &\cw  &\cw &\cw  &\cw  &\cw  &\cw &\cw  &\cw  &\cw  \\
		\lstick{\ket{A}}   &  &   &  &  &  &  &  &  &  &  &  &  &  &  &  &  &  &  &  &  &  &  &  &  &  &  &  & & & \ctrl{1} & \qw \qwx  &\cw  &\cw  &\cw  &\cw &\cw  &\cw  &\cw  &\cw &\cw  &\cw  &\cw  &\cw  &\cw  &\cw   \\
		\lstick{\ket{A}}   &  & \qw & \qw & \qw &\qw &\qw & \ctrl{3} &\qw  & \ctrl{3} &\qw & \qw & \qw & \qw & \qw & \qw & \qw & \qw & \qw & \qw & \qw & \targ & \qw & \qw & \qw &  \ctrl{2} & \qw &\qw & \qw  & \qw & \targ & \qw & \qw & \qw & \qw & \qw & \qw & \qw & \ctrl{4} &\qw  & \ctrl{4} &\qw  &\qw &\qw \qwx &\cw &\cw \\
		\lstick{\ket{a_3}} & \ctrl{5} & \ctrl{1} & \qw & \qw & \qw &\qw & \qw & \qw & \qw & \qw & \qw &\qw & \qw & \qw & \qw & \qw & \qw & \qw & \qw & \qw & \qw & \qw & \qw & \qw & \qw & \qw & \ctrl{1} & \qw & \qw & \qw & \qw & \qw & \qw & \qw & \qw & \qw & \qw & \qw & \qw & \qw &\qw & \ctrl{1} &\ctrl{2} & \qw & \qw &\rstick{\ket{a_3}} \\
		\lstick{\ket{b_3}} & \ctrl{5} & \targ & \ctrl{-5} & \qw &\qw & \qw & \ctrl{1} & \qw & \ctrl{1} & \qw & \qw & \qw & \qw & \qw & \qw & \qw & \qw & \qw & \qw & \qw & \qw &\qw &\qw  &\ctrl{-5} &\targ & \targ & \targ & \ctrl{-5} & \qw & \qw & \qw & \qw & \qw & \qw & \qw & \qw & \qw & \ctrl{4} & \qw & \ctrl{4} & \ctrl{-5} &\targ &\ctrl{5} & \targ & \qw &\rstick{\ket{s_3}} \\ 
		\lstick{\ket{A}}   &  & & & & & & & \ctrl{4} & \qw \qwx &\cw  &\cw  &\cw  &\cw &\cw  &\cw  &\cw  &\cw &\cw  &\cw  &\cw  &\cw &\cw  &\cw  &\cw  &\cw &\cw  &\cw  &\cw  &\cw &\cw  &\cw  &\cw  &\cw &\cw  &\cw  &\cw  &\cw &\cw  &\cw  &\cw  &\cw &\cw  &\cw  &\cw  &\cw   \\
		\lstick{\ket{A}}   &  & & & & & & &  &  &  &\ctrl{3}  &\qw \qwx &\cw  &\cw  &\cw  &\cw &\cw  &\cw  &\cw  &\cw &\cw  &\cw  &\cw  &\cw &\cw  &\cw  &\cw  &\cw &\cw  &\cw  &\cw  &\cw &\cw  &\cw  &\cw  &\cw &\cw  &\cw  &\cw  &\cw &\cw  &\cw  &\cw  &\cw &\cw  \\
		\lstick{\ket{A}}   &  & & & & & & &  &  &  &  & & &  &  &  &  &  &  &  &   &  &  &  &  &  &  &  & &  &  &  &  &  & &\ctrl{2} &\qw \qwx &\cw  &\cw  &\cw  &\cw &\cw  &\cw  &\cw  &\cw \\
		\lstick{\ket{A}}   &  & & & & & & &  &  &  &  & & &  &  &  &  &  &  &  &   &  &  &  &  &  &  &  & &  &  &  &  &  & & & & &\ctrl{1} &\qw \qwx &\cw  &\cw  &\cw  &\cw &\cw \\
		\lstick{\ket{A}}   &  & \qw & \qw & \qw & \qw & \qw &\qw & \targ &\qw  &\qw  & \targ & \qw & \ctrl{9} & \qw & \ctrl{9} & \ctrl{4} &\qw &\ctrl{4} & \qw & \ctrl{2} &\qw &\ctrl{2} & \qw & \qw &  \ctrl{2} & \qw & \qw & \qw & \ctrl{4} &\qw &\ctrl{4} & \ctrl{6} &\qw &\ctrl{6} & \qw  &\targ & \qw & \qw &\targ & \qw & \qw & \qw &\qw \qwx &\cw &\cw  \\
		\lstick{\ket{a_4}} & \ctrl{4} & \ctrl{1} & \qw &\qw & \qw & \qw & \qw & \qw & \qw & \qw & \qw & \qw & \qw &\qw & \qw & \qw & \qw & \qw & \qw & \qw & \qw & \qw & \qw & \qw & \qw & \qw &\ctrl{1} & \qw & \qw & \qw & \qw & \qw & \qw & \qw & \qw & \qw & \qw & \qw & \qw & \qw &\qw & \ctrl{1} &\ctrl{4} & \qw & \qw &\rstick{\ket{a_4}} \\
		\lstick{\ket{b_4}} & \ctrl{4} & \targ & \ctrl{2} & \qw & \qw & \qw & \qw & \qw & \qw &\qw & \qw &\qw & \qw &  \qw &\qw & \qw & \qw & \qw & \qw & \ctrl{2} &\qw &\ctrl{2} & \qw & \ctrl{4} & \targ & \targ & \targ & \ctrl{3} & \ctrl{3} & \qw & \ctrl{3} & \qw & \qw & \qw  & \qw & \qw & \qw & \qw & \qw & \qw & \ctrl{4} & \targ &\ctrl{4} & \targ & \qw &\rstick{\ket{s_4}} \\
		\lstick{\ket{A}}   &  &  &  & \qw & \ctrl{8} & \qw & \qw & \qw &  \qw & \qw & \qw & \qw & \qw &\qw & \qw & \ctrl{8} &\qw & \ctrl{8} & \qw  &  \qw & \qw & \qw & \ctrl{8} & \qw & \cw & \cw & \cw & \qw & \qw & \qw & \qw  &\ctrl{9} &\qw &\ctrl{9} & \qw & \qw &  \qw & \qw & \qw & \qw & \qw & \qw & \qw & \cw & \cw  \\
		\lstick{\ket{A}}   &  &  &  &  &  & & & &  & & & &  & & & &  & & & & \ctrl{2} & \qw \qwx & \cw & \cw & \cw & \cw & \cw & \cw & \cw & \cw & \cw & \cw & \cw & \cw & \cw & \cw & \cw & \cw & \cw & \cw & \cw & \cw & \cw & \cw & \cw \\
		\lstick{\ket{A}}   &  &  &  &  &  & & & &  & & & &  & & & &  & & & &  &  & &   &  &  &  &  &  &  \ctrl{1} & \qw \qwx & \cw & \cw & \cw & \cw & \cw & \cw & \cw & \cw & \cw & \cw & \cw & \cw & \cw  & \cw  \\
		\lstick{\ket{A}}   &  & \qw & \qw & \qw & \qw & \qw & \ctrl{3} & \qw &  \ctrl{3} &\qw & \qw & \qw & \qw & \qw &\qw & \qw &\qw & \qw & \qw & \qw &\targ  & \qw &\qw & \qw &  \ctrl{2} & \qw & \qw & \qw & \qw & \targ & \qw & \qw & \qw & \qw & \qw & \qw & \qw & \ctrl{4} & \qw & \ctrl{4} & \qw & \qw & \qw \qwx &\cw &\cw \\
		\lstick{\ket{a_5}} & \ctrl{5} & \ctrl{1} & \qw & \qw & \qw & \qw & \qw &  \qw &  \qw & \qw & \qw & \qw &\qw & \qw & \qw & \qw &\qw & \qw & \qw &  \qw & \qw & \qw & \qw & \qw & \qw & \qw &\ctrl{1} & \qw & \qw & \qw & \qw & \qw & \qw & \qw & \qw & \qw & \qw & \qw & \qw & \qw & \qw &\ctrl{1} &\ctrl{4} & \qw & \qw &\rstick{\ket{a_5}} \\
		\lstick{\ket{b_5}} & \ctrl{6} & \targ & \ctrl{-5} & \qw & \qw & \qw & \ctrl{2} &  \qw & \ctrl{2} & \qw & \qw & \qw &\qw & \qw & \qw & \qw & \qw & \qw &\qw &  \qw & \qw & \qw & \qw & \ctrl{-5} & \targ & \targ & \targ & \ctrl{-5} &\qw & \qw & \qw & \qw & \qw & \qw & \qw & \qw & \qw & \ctrl{5} & \qw & \ctrl{5} & \ctrl{-5} & \targ &\ctrl{5} & \targ & \qw &\rstick{\ket{s_5}} \\
		\lstick{\ket{A}}   &  &  &  &  &  & \qw & \qw & \qw & \qw & \qw & \qw & \qw & \ctrl{16} & \qw & \ctrl{16} & \qw & \qw & \qw &\qw & \qw & \qw &  \qw & \qw & \cw & \cw & \cw & \cw & \cw & \cw & \cw & \cw  & \cw & \cw & \cw & \cw & \cw & \cw & \cw & \cw & \cw & \cw & \cw & \cw  & \cw & \cw     \\
		\lstick{\ket{A}}   &  &  &  &  &  & & & \ctrl{4} & \qw \qwx & \cw & \cw & \cw & \cw  & \cw & \cw & \cw & \cw  & \cw & \cw & \cw & \cw  & \cw & \cw & \cw & \cw  & \cw & \cw & \cw & \cw  & \cw & \cw & \cw & \cw  & \cw & \cw & \cw & \cw  & \cw & \cw & \cw & \cw  & \cw & \cw & \cw &\cw \\
		\lstick{\ket{A}}   &  &  &  &  &  & & &  & & &  &  &  &  &  & &  \ctrl{3} & \qw \qwx & \cw & \cw & \cw & \cw  & \cw & \cw & \cw & \cw  & \cw & \cw & \cw & \cw  & \cw & \cw & \cw & \cw  & \cw & \cw & \cw & \cw  & \cw & \cw & \cw & \cw & \cw & \cw & \cw  \\ 
		\lstick{\ket{A}}   &  &  &  &  &  & & &  &  & &  &  &  &  &  & &  &  & &  &  & &  &  &  &  &  &  &  &  &  &  & \ctrl{2} & \qw \qwx  & \cw & \cw & \cw & \cw  & \cw & \cw & \cw & \cw  & \cw & \cw &\cw \\
		\lstick{\ket{A}}   &  &  &  &  &  & & &  &  & &  &  &  &  &  & &   &  & &  &  & &  &  &  &  &  &  &  &  &  &  &  &   & &  &  & & \ctrl{1} & \qw \qwx & \cw & \cw & \cw & \cw &\cw  \\
		\lstick{\ket{A}}   &  & \qw & \qw & \qw & \qw &\qw & \qw & \targ& \qw & \ctrl{5} & \qw & \ctrl{4} & \qw & \qw & \qw &\qw &\targ & \qw & \qw &  \ctrl{3} &\qw & \ctrl{3} &\qw & \qw &  \ctrl{2} & \qw & \qw & \qw &  \ctrl{3} &\qw & \ctrl{3} & \qw & \targ & \qw & \qw & \qw & \qw & \qw &\targ &  \qw & \qw &\qw &\qw \qwx & \cw & \cw \\
		\lstick{\ket{a_6}} & \ctrl{3} & \ctrl{1} & \qw & \qw & \qw & \qw & \qw &  \qw & \qw & \qw & \qw & \qw & \qw & \qw  &\qw & \qw & \qw &  \qw & \qw & \qw & \qw & \qw & \qw & \qw & \qw & \qw & \ctrl{1} & \qw & \qw &  \qw & \qw & \qw & \qw & \qw &  \qw & \qw & \qw & \qw & \qw & \qw & \qw &\ctrl{1} &\ctrl{4} & \qw & \qw &\rstick{\ket{a_6}} \\
		\lstick{\ket{b_6}} & \ctrl{4} & \targ & \ctrl{5} & \qw & \qw & \qw & \qw &  \qw & \qw & \qw &\qw & \qw &  \qw &\qw & \qw &\qw & \qw & \qw &  \qw  &\ctrl{2} & \qw & \ctrl{2} & \qw & \ctrl{3} & \targ & \targ & \targ & \qw & \ctrl{3} & \qw & \ctrl{3} & \qw & \qw & \qw & \qw &  \qw & \qw &  \qw & \qw & \qw & \qw & \targ &\ctrl{4} & \targ & \qw &\rstick{\ket{s_6}} \\
		\lstick{\ket{A}}   &  &  &  & \qw & \ctrl{-7} & \qw & \qw  &\qw &\qw & \ctrl{7} &  \qw & \ctrl{7} &\qw & \qw &\qw & \qw &\qw & \qw & \qw &  \qw & \qw &\qw & \ctrl{-9} &\qw  & \cw & \cw & \cw & \cw  & \cw & \cw & \cw & \cw  & \cw & \cw & \cw & \cw  & \cw & \cw & \cw & \cw  & \cw & \cw & \cw & \cw  & \cw  \\
		\lstick{\ket{A}}   &  &   &  &  &  &  &  &  &  &  &  &  &  &  &  &  &  &  &  &  &\ctrl{2} &\qw \qwx & \cw & \cw & \cw & \cw  & \cw & \cw & \cw & \cw  & \cw & \cw & \cw & \cw  & \cw & \cw & \cw & \cw  & \cw & \cw & \cw & \cw  & \cw & \cw  & \cw \\
		\lstick{\ket{A}}   &  &   &  &  &  &  &  &  &  &  &  &  &  &  &  &  &  &  &  &  & & &  &  &  &  &  &  &  &\ctrl{1} &\qw \qwx & \cw & \cw & \cw & \cw  & \cw & \cw & \cw & \cw  & \cw & \cw & \cw & \cw & \cw & \cw\\
		\lstick{\ket{A}}   &  & \qw & \qw & \qw & \qw & \qw & \ctrl{2}  & \qw &  \ctrl{3} & \qw & \qw & \qw & \qw &\qw & \qw & \qw & \qw & \qw &  \qw & \qw & \targ & \qw & \qw & \qw &  \ctrl{2} & \qw & \qw & \qw & \qw &\targ & \qw & \qw & \qw & \qw & \qw & \qw & \qw & \qw & \qw & \qw & \qw & \qw &\qw \qwx& \cw & \cw   \\
		\lstick{\ket{a_7}} & \ctrl{4} & \ctrl{1} & \qw & \qw & \qw & \qw & \qw & \qw &  \qw & \qw & \qw & \qw & \qw & \qw &\qw & \qw & \qw &\qw & \qw & \qw & \qw & \qw & \qw & \qw & \qw & \qw & \qw & \qw & \qw & \qw & \qw & \qw & \qw & \qw & \qw & \qw & \qw & \qw & \qw & \qw & \qw & \qw & \qw & \qw & \qw&\rstick{\ket{a_7}}  \\
		\lstick{\ket{b_7}} & \ctrl{4} & \targ & \ctrl{-5} & \qw & \qw & \qw & \ctrl{1} & \qw &  \ctrl{1} & \qw & \qw &\qw & \qw & \qw &\qw & \qw & \qw & \qw & \qw & \qw & \qw & \qw & \qw & \ctrl{-5} & \targ & \qw & \qw & \qw & \qw & \qw & \qw & \qw & \qw & \qw & \qw & \qw & \qw & \qw & \qw & \qw & \qw & \qw & \qw & \qw & \qw&\rstick{\ket{s_7}} \\
		\lstick{\ket{A}}   & &  &  &  &  & & & \ctrl{3} & \qw \qwx & \cw & \cw & \cw & \cw & \cw & \cw & \cw & \cw & \cw & \cw & \cw & \cw & \cw & \cw & \cw & \cw & \cw & \cw & \cw & \cw & \cw & \cw & \cw & \cw & \cw & \cw & \cw & \cw & \cw & \cw & \cw & \cw & \cw & \cw & \cw & \cw  \\
		\lstick{\ket{A}}   & &  &  &  &  & & &  &  & &\ctrl{2} & \qw \qwx & \cw & \cw & \cw & \cw & \cw & \cw & \cw & \cw & \cw & \cw & \cw & \cw & \cw & \cw & \cw & \cw & \cw & \cw & \cw & \cw & \cw & \cw & \cw & \cw & \cw & \cw & \cw & \cw & \cw & \cw & \cw & \cw & \cw \\
		\lstick{\ket{A}}   & &  &  &  &  & & &  &  & & &  &  & \ctrl{1} & \qw \qwx & \cw & \cw & \cw & \cw & \cw & \cw & \cw & \cw & \cw & \cw & \cw & \cw & \cw & \cw & \cw & \cw & \cw & \cw & \cw & \cw & \cw & \cw & \cw & \cw & \cw & \cw & \cw & \cw & \cw & \cw \\
		\lstick{\ket{A}}   &  & \qw & \qw &\qw & \qw & \qw &\qw &\targ &\qw & \qw & \targ & \qw & \qw & \targ & \qw & \qw & \qw & \qw & \qw & \qw &\qw & \qw &\qw & \qw &\qw & \qw & \qw & \qw & \qw & \qw & \qw & \qw & \qw & \qw & \qw & \qw & \qw & \qw &\qw & \qw & \qw & \qw & \qw & \qw & \qw &\rstick{\ket{s_8}} \\
	}
	\]
	\caption{In-place FT-QCLA1 (In-FT-QCLA1) for the case of adding two $8$ bit values $a$ and $b$.}
	\label{IP-TgateOptimized}
\end{figure}

The T gate optimized fault tolerant QCLA is targeted for quantum hardware that can support the resource requirements needed for fault tolerant quantum computation.  Because of the high implementation cost of the fault tolerant T gate, this QCLA is optimized for T-count.  The proposed QCLA is based on the quantum NOT gate, CNOT gate, logical AND gate and the uncomputation gate presented in \cite{Gidney20184TgateToffolibuild}.  The steps of the methodology to implement the proposed In-FT-QCLA1 are shown along with an illustrative example of the QCLA circuit in Figure \ref{IP-TgateOptimized}.

\begin{itemize}
	
	\item \textbf{Step 1:} For \textit{i} = 0 to n-1, apply the logical AND gate at locations A[\textit{i}], B[\textit{i}] and an ancillae such that locations A[\textit{i}] and B[\textit{i}] are unchanged.  The ancillae will have the result of computation.  The ancillae will be renamed to the value g[\textit{i}, \textit{i} +1].  
	
	\item \textbf{Step 2:} For \textit{i} = 1 to n-1: At locations of A[\textit{i}] and B[\textit{i}] apply a CNOT gate such that location A[\textit{i}] is unchanged and location B[\textit{i}] will have the result of computation.  Location B[\textit{i}] is renamed to the value p[\textit{i}, \textit{i}+1].     
	
	\item \textbf{Step 3 (P-rounds):}  We use a nested loop in this step.  For $t = 1 \text{ to } \lfloor log(n) \rfloor-1$ and For $m = 1 \text{ to } \lfloor \frac{n}{2^t} \rfloor -1$:
	
	At locations p[\textit{j},\textit{l}], p[\textit{l}, \textit{k}] and an ancillae apply a logical AND gate such that locations p[\textit{j},\textit{l}] and p[\textit{l}, \textit{k}] are unchanged.  The ancillae will have the result of computation and will be renamed to the value p[\textit{j},\textit{k}].  The equations for the indexes are $j = 2^t \cdot m$, $k = 2^t \cdot m + 2^t$ and $l = 2^t \cdot m + 2^{t-1}$. 
	
	\item \textbf{Step 4 (G-rounds):}  We use a nested loop in this step.  For $t = 1 \text{ to } \lfloor log(n) \rfloor$ and For $m = 0 \text{ to } \lfloor \frac{n}{2^t} \rfloor -1$:
	
	At locations g[\textit{j}, \textit{l}], p[\textit{l}, \textit{k}], and g[\textit{l}, \textit{k}] apply a Toffoli gate such that locations g[\textit{j}, \textit{l}] and p[\textit{l}, \textit{k}] pass through unchanged.  Location g[\textit{l}, \textit{k}] holds the result of computation and will be renamed to the value g[\textit{j}, \textit{k}].  The equations for the indexes are $j = 2^t \cdot m$, $k = 2^t \cdot m + 2^t$, and $l = 2^t \cdot m + 2^{t-1}$.
	
	\item \textbf{Step 5 (C-rounds):}  We use a nested loop in this step.  For $t = \lfloor log \left( \frac{2n}{3} \right) \rfloor \text{ to } 1$ and For $m =  1 \text{ to } \lfloor \frac{n - 2^{t-1}}{2^t} \rfloor$: 
	
	At locations g[0, \textit{l}], p[\textit{l},\textit{k}], and g[\textit{l},\textit{k}] apply a Toffoli gate such that locations g[0, \textit{l}] and p[\textit{l},\textit{k}] are unchanged.   Location g[\textit{l},\textit{k}] will hold the result of computation and will be renamed to the value g[0, \textit{k}].  The equations for the indexes are $l = 2^t \cdot m$ and $k = 2^t \cdot m + 2^{t-1}$.   
	
	\item \textbf{Step 6 (P-erase-rounds):}  We use a nested loop in this step.  For $t = \lfloor log(n) \rfloor - 1 \text{ to } 1$ and For $m =  1 \text{ to } \lfloor \frac{n}{2^t} \rfloor - 1$:
	
	At locations p[\textit{j}, \textit{l}], p[\textit{l}, \textit{k}] and p[\textit{j}, \textit{k}] apply a uncomputation gate such that locations p[\textit{j}, \textit{l}] and p[\textit{l}, \textit{k}] are unchanged.  Location p[\textit{j}, \textit{k}] will be restored to its original value.  The equation for the indexes are $j = 2^t \cdot m$, $k = 2^t \cdot m + 2^t$ and $l = 2^t \cdot m + 2^{t-1}$.    
	
	\item \textbf{Step 7:} For \textit{i}  = 1 to n-1: at locations p[\textit{i}, \textit{i} +1] and g[0, \textit{i}] apply a CNOT gate such that location g[0, \textit{i}] is unchanged.  Location p[\textit{i}, \textit{i} +1] will have the result of computation.
	
	\item \textbf{Step 8:} This step has the following two sub-steps:
	
	\begin{itemize}
		
		\item \textbf{Sub-step 1:} At location B[\textit{0}] apply a NOT gate.  The location B[\textit{0}] will be renamed to the value p[\textit{0}, 1]
		
		\item \textbf{Sub-step 2:} For \textit{i} = 1 to n-2: At location p[\textit{i}, \textit{i} +1] apply a NOT gate.
		
	\end{itemize}
	
	\item \textbf{Step 9:} For \textit{i} = 1 to n-2: At locations A[\textit{i}] and p[\textit{i}, \textit{i} +1] apply the CNOT gate such that location A[\textit{i}] is unchanged and p[\textit{i}, \textit{i} +1] has the result of computation.
	
	\item \textbf{Step 10 (Reverse of P-erase-rounds):} We use a nested loop in this step.  For $t = 1 \text{ to } \lfloor log(n) \rfloor - 1$ and For $m = 1 \text{ to } \lfloor \frac{n}{2^t} \rfloor - 1$:
	
	At locations p[\textit{j},\textit{l}], p[\textit{l},\textit{k}] and an ancillae apply a logical AND gate such that locations p[\textit{j},\textit{l}] and p[\textit{l},\textit{k}] are unchanged.  The ancillae will hold the result of computation and the ancillae is renamed to the value p[\textit{j},\textit{k}].  The equations for the indexes are $j = 2^t \cdot m$, $k = 2^t \cdot m + 2^t$, and $l = 2^t \cdot m + 2^{t-1}$. %[ARE INDICES CORRECT!!!]      
	
	\item \textbf{Step 11 (Reverse of C-rounds):} We use a nested loop in this step.  For $t = 1 \text{ to } \lfloor log \left( \frac{2n}{3} \right) \rfloor$ and For $m =  1 \text{ to } \lfloor \frac{(n-2^{t-1})}{2^t} \rfloor$:
	
	At locations g[0,\textit{l}], p[\textit{l},\textit{k}], and g[0,\textit{k}] apply a Toffoli gate such that locations g[0,\textit{l}] and p[\textit{l},\textit{k}] are unchanged.  Location g[0,\textit{k}] will have the result of computation and the location will be renamed to the value g[\textit{l},\textit{k}].  The equations for the indexes are $l = 2^t \cdot m$ and $k = 2^t \cdot m + 2^{t-1}$. %[ARE INDICES CORRECT!!!]
	
	\item \textbf{Step 12 (Reverse of G-rounds):} We use a nested loop in this step. For $t = \lfloor log(n) \rfloor \text{ to } 1$ and For $ m =  0 \text{ to } \lfloor \frac{n}{2^t} \rfloor - 1$:
	
	At locations g[\textit{j},\textit{l}], p[\textit{l},\textit{k}] and g[\textit{j},\textit{k}] apply a Toffoli gate such that locations g[\textit{j},\textit{l}] and p[\textit{l},\textit{k}] are unchanged.  Location g[\textit{j},\textit{k}] will have the result of computation and the location is renamed to the value g[\textit{l},\textit{k}. The equations for the indexes are $j = 2^t \cdot m$, $k = 2^t \cdot m + 2^t$ and $l = 2^t \cdot m + 2^{t-1}$. %[ARE INDICES CORRECT!!!]
	
	\item \textbf{Step 13 (Reverse of P-rounds):} We use a nested loop in this step. For $t = \lfloor log(n) \rfloor - 1 \text{ to } 1$ and For $m = 1  \text{ to } \lfloor \frac{n-1}{2^t} \rfloor - 1$:
	
	At locations p[\textit{j},\textit{l}], p[\textit{j},\textit{k}], and p[\textit{l},\textit{k}] apply a uncomputation gate such that locations p[\textit{j},\textit{l}] and p[\textit{l},\textit{k}] are unchanged.  Location p[\textit{j},\textit{k}] will be restored to its original value.  The equations for the indexes are $j = 2^t \cdot m$, $k = 2^t \cdot m + 2^t$ and $l = 2^t \cdot m + 2^{t-1}$  %[NO INDICES DEFINITION !!!!] [ARE GIVEN INDICES CORRECT!!!]
	
	\item \textbf{Step 14:} For i=1 to n-2: At locations A[\textit{i}] and p[\textit{i}, \textit{i} +1], apply a CNOT gate such that location A[\textit{i}] would be unchanged and location p[\textit{i}, \textit{i} +1] will have the result of computation.
	
	\item \textbf{Step 15:} For i=0 to n-2: at locations A[i], p[\textit{i}, \textit{i} +1], and g[\textit{i}, \textit{i} +1], apply an uncomputation gate such that locations a[i] and p[\textit{i}, \textit{i} +1] are unchanged and g[\textit{i}, \textit{i} +1] will be restored to its original value.     
	
	\item \textbf{Step 16:} Location p[\textit{n-1}, \textit{n}] has the sum bit $s_{{\textit{n-1}}}$ and g[\textit{0}, \textit{n}] has the sum bit $s_{\textit{n}}$.  For i=0 to n-2: At location p[\textit{i}, \textit{i} +1] apply a NOT gate.  Location p[\textit{i}, \textit{i} +1] will have the sum bit $s_{{\textit{i}}}$. 
	
	%$^*$To preserve the last carry value or $s_{{\textit{n}}}$, gates and circuits in steps 10 to 13 that do an operation on the last carry value needs to be deleted so they would not interfere with the last carry value. Therefore the bounds for t and m need to be adjusted.
	
\end{itemize}

\subsection{Proposed In-Place FT-QCLA2 Design (In-FT-QCLA2)}
\label{IP-fault-tolerant-QCLA}

In-FT-QCLA1 is optimized for T-count.  A trade-off for the reduced T-count of In-FT-QCLA1 is an increase in the qubit cost of In-FT-QCLA1.  We can reduce the qubit cost by replacing logical-AND gate and uncomputation gate pairs with alternative Toffoli gates implementations (such as the design in \cite{Maslov}) where appropriate.  We save qubits because the logical AND gate and uncomputation gate pairs have a qubit cost of $4$ while implementations like the design in \cite{Maslov} have a qubit cost of $3$.  However, alternative Toffoli gate implementations have higher T-counts (such as the design in \cite{Maslov} with a T-count of $7$).

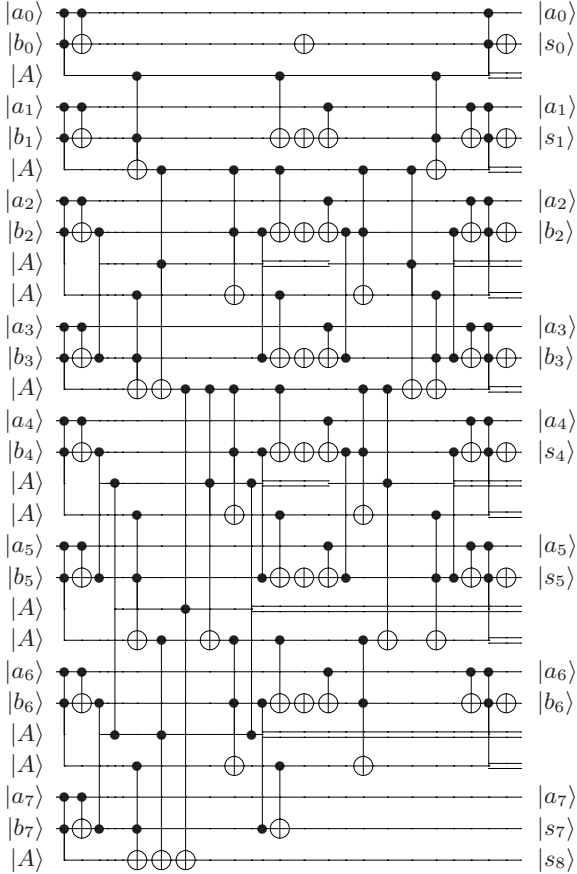
\begin{figure}[htbp]
	\centering
	\small
	\[
	\Qcircuit @C=0.2em @R=0.5em @!R{
		\lstick{\ket{a_0}} & \ctrl{2} & \ctrl{1} & \qw & \qw & \qw & \qw & \qw & \qw & \qw & \qw & \qw & \qw & \qw & \qw & \qw & \qw &\qw & \qw & \qw & \qw & \qw & \qw & \qw & \ctrl{2} & \qw & \qw & \rstick{\ket{a_0}} \\
		\lstick{\ket{b_0}} & \ctrl{1} & \targ & \qw & \qw & \qw & \qw & \qw & \qw & \qw & \qw & \qw & \qw & \qw & \qw & \targ & \qw &\qw & \qw & \qw & \qw & \qw & \qw & \qw &  \ctrl{1} & \targ & \qw & \rstick{\ket{s_0}}\\
		\lstick{\ket{A}}   &  & \qw & \qw &\qw & \qw & \qw & \ctrl{3} & \qw & \qw & \qw & \qw & \qw & \qw & \ctrl{2} & \qw & \qw &\qw & \qw & \qw & \qw & \ctrl{2} & \qw &  \qw & \qw &\cw  &\cw  &\\
		\lstick{\ket{a_1}} & \ctrl{2} & \ctrl{1} & \qw & \qw & \qw & \qw & \qw & \qw & \qw & \qw & \qw & \qw & \qw & \qw & \qw & \ctrl{1} &\qw & \qw & \qw & \qw & \qw & \qw & \ctrl{1} & \ctrl{2} & \qw &\qw & \rstick{\ket{a_1}}\\ 
		\lstick{\ket{b_1}} & \ctrl{1} & \targ & \qw & \qw & \qw & \qw & \ctrl{1} & \qw & \qw & \qw & \qw & \qw &\qw & \targ & \targ & \targ & \qw & \qw & \qw & \qw & \ctrl{1} & \qw & \targ &  \ctrl{1} & \targ &\qw & \rstick{\ket{s_1}}\\ 
		\lstick{\ket{A}}   & & \qw & \qw & \qw & \qw & \qw & \targ & \ctrl{4} & \qw & \qw & \ctrl{3} & \qw & \qw & \ctrl{2} & \qw & \qw & \qw & \ctrl{2} & \qw & \ctrl{4} & \targ & \qw &  \qw & \qw &\cw  &\cw  &\\  
		\lstick{\ket{a_2}} & \ctrl{2} & \ctrl{1} & \qw & \qw & \qw & \qw & \qw & \qw & \qw & \qw & \qw & \qw & \qw & \qw & \qw & \ctrl{1} & \qw & \qw & \qw & \qw & \qw & \qw &  \ctrl{1} & \ctrl{2} &  \qw &\qw & \rstick{\ket{a_2}} \\
		\lstick{\ket{b_2}} & \ctrl{2} & \targ & \ctrl{2} & \qw &\qw & \qw & \qw & \qw & \qw & \qw & \ctrl{2} & \qw & \ctrl{2} & \targ &\targ & \targ & \ctrl{2} & \ctrl{2} & \qw & \qw & \qw & \ctrl{2} & \targ &  \ctrl{2} & \targ &\qw & \rstick{\ket{b_2}}\\
		\lstick{\ket{A}}   &  &  &  & \qw &\qw & \qw & \qw & \ctrl{4} & \qw & \qw & \qw &\qw &\qw  &\cw  &\cw  &\cw  &\qw  &\qw & \qw & \ctrl{4} & \qw & \qw &\cw   &\cw  &\cw  &\cw  &\\
		\lstick{\ket{A}}   &  & \qw & \qw & \qw &\qw & \qw & \ctrl{3} & \qw & \qw & \qw & \targ &\qw & \qw & \ctrl{2} & \qw & \qw & \qw & \targ & \qw & \qw & \ctrl{2} & \qw & \qw & \qw &\cw  &\cw  &\\
		\lstick{\ket{a_3}} & \ctrl{2} & \ctrl{1} & \qw & \qw &\qw & \qw & \qw & \qw & \qw & \qw & \qw &\qw & \qw & \qw & \qw & \ctrl{1} &\qw & \qw & \qw & \qw & \qw & \qw &  \ctrl{1} & \ctrl{2} & \qw &\qw & \rstick{\ket{a_3}}\\
		\lstick{\ket{b_3}} & \ctrl{1} & \targ & \ctrl{-2} & \qw &\qw & \qw & \ctrl{1} & \qw & \qw & \qw & \qw & \qw & \ctrl{-2} & \targ & \targ & \targ & \ctrl{-2} &\qw & \qw & \qw & \ctrl{1} & \ctrl{-2} & \targ &  \ctrl{1} & \targ & \qw & \rstick{\ket{b_3}}\\  
		\lstick{\ket{A}}   &  & \qw & \qw & \qw & \qw & \qw & \targ & \targ & \ctrl{7} & \ctrl{4} &\ctrl{3} & \qw & \qw & \ctrl{2} & \qw & \qw & \qw & \ctrl{2} & \ctrl{3} & \targ & \targ &  \qw &  \qw & \qw &\cw  &\cw  &\\
		\lstick{\ket{a_4}} & \ctrl{2} & \ctrl{1} & \qw &\qw & \qw & \qw & \qw & \qw & \qw & \qw & \qw & \qw & \qw & \qw & \qw & \ctrl{1} & \qw &\qw & \qw & \qw & \qw  &  \qw &  \ctrl{1} & \ctrl{2} & \qw & \qw & \rstick{\ket{a_4}} \\
		\lstick{\ket{b_4}} & \ctrl{2} & \targ & \ctrl{2} & \qw & \qw & \qw & \qw & \qw & \qw &\qw & \ctrl{2} & \qw & \ctrl{2} & \targ & \targ & \targ & \ctrl{2} & \ctrl{2} & \qw & \qw & \qw & \ctrl{2} & \targ &  \ctrl{2} & \targ & \qw & \rstick{\ket{s_4}} \\
		\lstick{\ket{A}}   &  &  &  & \qw & \ctrl{4} & \qw & \qw & \qw &  \qw & \ctrl{5} & \qw &\ctrl{4} & \qw &\cw  &\cw &\cw  &\qw  &\qw &\ctrl{5} & \qw & \qw & \qw &\cw  &\cw  &\cw  &\cw  &\\
		\lstick{\ket{A}}   &  & \qw & \qw & \qw & \qw & \qw & \ctrl{3} & \qw &  \qw &\qw & \targ & \qw & \qw & \ctrl{2} & \qw & \qw & \qw & \targ & \qw & \qw & \ctrl{2} & \qw &  \qw & \qw &\cw  &\cw  &\\
		\lstick{\ket{a_5}} & \ctrl{2} & \ctrl{1} & \qw & \qw & \qw & \qw & \qw &  \qw &  \qw & \qw & \qw & \qw & \qw & \qw & \qw & \ctrl{1} & \qw &\qw & \qw & \qw & \qw & \qw &  \ctrl{1} & \ctrl{2} & \qw & \qw & \rstick{\ket{a_5}}\\
		\lstick{\ket{b_5}} & \ctrl{2} & \targ & \ctrl{-2} & \qw & \qw & \qw & \ctrl{2} &  \qw & \qw & \qw & \qw & \qw & \ctrl{-2} & \targ & \targ & \targ & \ctrl{-2} &\qw &\qw & \qw &\ctrl{2} & \ctrl{-2} & \targ &  \ctrl{2} & \targ & \qw & \rstick{\ket{s_5}}\\
		\lstick{\ket{A}}   &  &  &  &  &  & \qw & \qw & \qw &  \ctrl{8} & \qw & \qw & \qw &\cw  &\cw  &\cw  &\cw  &\cw  &\cw  &\cw  &\cw  &\cw  &\cw  &\cw  &\cw  &\cw  &\cw &\\
		\lstick{\ket{A}}   &  & \qw & \qw & \qw & \qw & \qw & \targ& \ctrl{3} & \qw & \targ & \ctrl{3} & \qw & \qw & \ctrl{2} & \qw & \qw & \qw & \ctrl{2} & \targ & \qw & \targ & \qw &  \qw & \qw &\cw  &\cw  &\\
		\lstick{\ket{a_6}} & \ctrl{2} & \ctrl{1} & \qw & \qw & \qw & \qw & \qw &  \qw & \qw & \qw & \qw & \qw & \qw & \qw & \qw & \ctrl{1} & \qw & \qw & \qw & \qw & \qw & \qw &  \ctrl{1} & \ctrl{2} & \qw & \qw & \rstick{\ket{a_6}} \\
		\lstick{\ket{b_6}} & \ctrl{2} & \targ & \ctrl{2} & \qw & \qw & \qw & \qw &  \qw & \qw & \qw & \ctrl{2} & \qw &\ctrl{2} &\targ & \targ & \targ & \qw & \ctrl{2} & \qw & \qw & \qw & \qw & \targ & \ctrl{2} & \targ & \qw & \rstick{\ket{b_6}}\\
		\lstick{\ket{A}}   &  &  &  & \qw & \ctrl{-4} & \qw & \qw  & \ctrl{4} &  \qw & \qw & \qw &\ctrl{-4} & \qw &\cw  &\cw  &\cw  &\cw  &\cw  &\cw  &\cw  &\cw  &\cw  &\cw  &\cw  &\cw  & \cw &\\
		\lstick{\ket{A}}   &  & \qw & \qw & \qw & \qw & \qw & \ctrl{2}  & \qw &  \qw & \qw & \targ & \qw & \qw & \ctrl{2} &\qw & \qw & \qw & \targ & \qw & \qw & \qw & \qw &  \qw & \qw &\cw  &\cw  &\\
		\lstick{\ket{a_7}} & \ctrl{2} & \ctrl{1} & \qw & \qw & \qw & \qw & \qw & \qw &  \qw & \qw & \qw & \qw & \qw & \qw & \qw & \qw & \qw & \qw & \qw & \qw & \qw & \qw &  \qw & \qw & \qw &\qw &  \rstick{\ket{a_7}}\\
		\lstick{\ket{b_7}} & \ctrl{1} & \targ & \ctrl{-2} & \qw & \qw & \qw & \ctrl{1} & \qw &  \qw & \qw & \qw & \qw & \ctrl{-2} & \targ & \qw & \qw & \qw & \qw & \qw & \qw &\qw & \qw &  \qw & \qw & \qw & \qw & \rstick{\ket{s_7}}\\
		\lstick{\ket{A}}   &  & \qw & \qw &\qw & \qw & \qw &\targ &\targ &\targ & \qw & \qw & \qw & \qw & \qw & \qw & \qw & \qw & \qw & \qw & \qw & \qw & \qw &  \qw & \qw & \qw &\qw & \rstick{\ket{s_8}} \\
	}
	\]
	\caption{In-place FT-QCLA2 (In-FT-QCLA2) for the case of adding two $8$ bit values $a$ and $b$.}
	\label{IP-QubitOptimized}
\end{figure}

The steps of the methodology to implement In-FT-QCLA2 are identical to the methodology to implement In-FT-QCLA1.  To implement In-FT-QCLA2, replace the logical AND gate, uncomputation gate pairs with a Toffoli gate in Step 4, Step 5, Step 11 and Step 12.  An illustrative example of the In-FT-QCLA2 circuit is in Figure \ref{IP-QubitOptimized}.

\section{Performance of Proposed In-place QCLA circuits}
\label{IP-QCLA-performance}

\begin{table*}[tbhp]
	
	\centering
	\caption{Cost Comparison of In-Place QCLAs}
	%\begin{tabular}{ |p{8cm}||p{8cm}|p{8cm}|p{8cm}| }
	\begin{tabular}{lcccc}
		\\ \midrule												
		Design	&	& T-count Equation	&	& Qubit Equation  \\		\cmidrule{5-5} \cmidrule{1-5}			
		Draper et al. (\cite{Draper2006qcla})	&	&	$70n-21w(n)-21\lfloor log(n)\rfloor-21w(n-1)-21\lfloor log (n-1) \rfloor-49$                      &	&	$4n-w(n)-\lfloor log(n)\rfloor+1$  	\\ 
		Trisetyarso et al. (\cite{Trisetyarso2009qcla}) &   &  $70n-21w(n)-21\lfloor log(n)\rfloor-21w(n-1)-21\lfloor log (n-1) \rfloor-49$  &   &   $4n-w(n)-\lfloor log(n)\rfloor+1$        \\
		Thapliyal et al. (\cite{Thapliyal2013qcla})	&   &   $\frac{203}{4}n-28$      &   &   $4n+1$            \\
		Takahashi et al.(\cite{Takahashi2008qcla}) & & $\approx 196n$ & & $\approx 5n$   \\
		Takahashi et al.(\cite{Takahashi2010inlaceQCLA}) & & $\approx 49n$ & & $\approx 5n$  \\
		Cheng et al. (\cite{Cheng2002QCLAinplace}) & & $\frac{14}{6} n^3 + \frac{21}{6} n^2 - \frac{49}{6} n$ & & $3 \cdot n + 1$  \\
		Mogenson (Design 1)$^1$ (\cite{Mogensen2019qcla}) & & $\approx 84 \cdot n - 56$ & & $3 \cdot n - 1$  \\
		Mogenson (Design 2)$^2$ (\cite{Mogensen2019qcla}) & & $\approx 84 \cdot n - 56$ & & $3 \cdot n - log(n) - 1$  \\
		In-FT-QCLA1$^3$  &   & $20n-8w(n)-8w(n-1)-4\lfloor log(n) \rfloor-4\lfloor log(n-1) \rfloor-8$  &   & $6n-2w(n)-2\lfloor log(n) \rfloor$  \\
		In-FT-QCLA2$^4$  &   & $40n - 11w(n) -  11\lfloor log(n) \rfloor - 11w(n-1) -  11\lfloor log(n-1) \rfloor - 32$  &   & $4n - w(n) -  \lfloor log(n) \rfloor +1$ \\ \bottomrule	
		\multicolumn{5}{l}{$w(n) = n - \sum_{y=1}^{\infty} \bigl\lfloor \frac{n}{2^y} \bigr\rfloor$} \\
		\multicolumn{5}{l}{$^1$ Mogenson (Design 1) can accept a carry in bit $c_0$.} \\
		\multicolumn{5}{l}{$^2$ Mogenson (Design 2) does not accept a carry in bit.} \\
		\multicolumn{5}{l}{$^3$ In-FT-QCLA1 is optimized emphasizing T-count.} \\
		\multicolumn{5}{l}{$^4$ In-FT-QCLA2 is optimized emphasizing qubit cost.} \\
	\end{tabular}
	\label{Equations for In-place T-count}
\end{table*}

\subsection{T-count analysis of In-FT-QCLA1}

The T-count of In-FT-QCLA1 is shown for each Step of the proposed design methodology.  Total T-count is determined by summing the T-count for each Step of the proposed design methodology.  The total T-count is  $20 \cdot n - 8 \cdot w(n) -  4 \cdot \lfloor log(n) \rfloor - 8 \cdot w(n-1) -  4 \cdot \lfloor log(n-1) \rfloor - 8$ where $w(n) = n - \sum_{y=1}^{\infty} \bigl\lfloor \frac{n}{2^y} \bigr\rfloor$.

\begin{itemize}
	\item Step 1 uses $n$ logical AND gates.  The T-count for this step is $4 \cdot n$.
	\item Step 2 does not need T gates. 
	\item Step 3 uses $n - w(n) - \lfloor log(n) \rfloor$ logical AND gates.  The T-count for this step is $4 \cdot (n - w(n) - \lfloor log(n) \rfloor)$. 
	\item Step 4 uses $n - w(n)$ Toffoli gates.  The T-count for this step is $4 \cdot (n - w(n))$. 
	\item Step 5 uses $n - \lfloor log(n) \rfloor - 1$ Toffoli gates.  The T-count for this step is $4 \cdot (n - \lfloor log(n) \rfloor - 1)$. 
	\item Steps 6 through 9 does not need T gates.
	\item Step 10 uses $n-1 - w(n-1) - \lfloor log(n-1) \rfloor$ logical AND gates.  The T-count for this step is $4 \cdot (n-1 - w(n-1) - \lfloor log(n-1) \rfloor)$.
	\item Step 11 uses $n- \lfloor log(n-1) \rfloor - 2$ Toffoli gates.  The T-count for this step is $4 \cdot (n- \lfloor log(n-1) \rfloor - 2)$.
	\item Step 12 uses $n-1 - w(n-1)$ Toffoli gates. The T-count for this step is $4 \cdot (n-1 - w(n-1))$.
	\item Steps 13 through 16 does not need T gates.
\end{itemize}

\subsection{T-count analysis of In-FT-QCLA2}

The T-count of In-FT-QCLA2 is shown for each Step of the proposed design methodology.  Total T-count is determined by summing the T-count for each Step of the proposed design methodology.  The total T-count is $40 \cdot n - 11 \cdot w(n) -  11 \cdot \lfloor log(n) \rfloor - 11 \cdot w(n-1) -  11 \cdot \lfloor log(n-1) \rfloor - 32$ where $w(n) = n - \sum_{y=1}^{\infty} \bigl\lfloor \frac{n}{2^y} \bigr\rfloor$.

\begin{itemize}
	\item Step 1 uses $n$ logical AND gates.  The T-count for this step is $4 \cdot n$.
	\item Step 2 does not need T gates. 
	\item Step 3 uses $n - w(n) - \lfloor log(n) \rfloor$ logical AND gates.  The T-count for this step is $4 \cdot (n - w(n) - \lfloor log(n) \rfloor)$. 
	\item Step 4 uses $n - w(n)$ Toffoli gates.  The T-count for this step is $7 \cdot (n - w(n))$. 
	\item Step 5 uses $n - \lfloor log(n) \rfloor - 1$ Toffoli gates.  The T-count for this step is $7 \cdot (n - \lfloor log(n) \rfloor - 1)$. 
	\item Steps 6 through 9 does not need T gates.
	\item Step 10 uses $n-1 - w(n-1) - \lfloor log(n-1) \rfloor$ logical AND gates.  The T-count for this step is $4 \cdot (n-1 - w(n-1) - \lfloor log(n-1) \rfloor)$.
	\item Step 11 uses $n- \lfloor log(n-1) \rfloor - 2$ Toffoli gates.  The T-count for this step is $7 \cdot (n- \lfloor log(n-1) \rfloor - 2)$.
	\item Step 12 uses $n-1 - w(n-1)$ Toffoli gates. The T-count for this step is $7 \cdot (n-1 - w(n-1))$.
	\item Steps 13 through 16 does not need T gates.
\end{itemize}

\subsection{Cost Comparison of Proposed In-Place QCLAs}

\subsubsection{Cost Comparison in Terms of T-count}		

Table \ref{Equations for In-place T-count} indicates that all proposed in-place QCLAs have T-count costs of order $\mathcal{O}(n)$.  The existing works also have a T-count of order $\mathcal{O}(n)$ with the exception of Cheng et al. where the T-count is of order $\mathcal{O}(n^3)$.  Of the proposed designs, In-FT-QCLA1 has the lowest T-count.  

%%%%%%%%%%%%%%%%%%%%%%%%%%%%%%%%%%%%%%%%%%%%%%%%%%%%%%%%%%%%%%%%%%%%%%%%%%%%%%%%%%%%%%%%%%%%%%%%%%%%%%%%%%%%

In-FT-QCLA2 requires roughly $79.59 \%$ fewer T gates than the design by Takahashi et al. in \cite{Takahashi2008qcla}, $18.37 \%$ fewer T gates than the design by Takahashi et al. in \cite{Takahashi2010inlaceQCLA}, $52.38 \%$ fewer T gates than the designs by Mogenson and $42.86 \%$ fewer T gates than the designs by Draper et al. and Trisetyarso et al. and $21.18 \%$ fewer T gates than the Thapliyal et al.  

The Proposed In-place In-FT-QCLA1 requires roughly $89.80 \%$ fewer T gates than the design by Takahashi et al. in \cite{Takahashi2008qcla}, $59.18 \%$ fewer T gates than the design by Takahashi et al. in \cite{Takahashi2010inlaceQCLA}, $76.19 \%$ fewer T gates than the design by Mogenson and $71.43 \%$ fewer T gates than the designs by Draper et al. and Trisetyarso et al. and $60.59 \%$ fewer T gates compared to the design by Thapliyal et al.  The In-FT-QCLA1 and In-FT-QCLA2 also achieve an order of magnitude reduction in T gates compared to the design by Cheng et al. 

\subsubsection{Cost Comparison in Terms of Qubits}	

Table \ref{Equations for In-place T-count} indicates that the proposed in-place QCLAs have qubit costs of order $\mathcal{O}(n)$.  The existing works also have a T-count of order $\mathcal{O}(n)$.  In-FT-QCLA2 requires $w(n)+\lfloor log(n)\rfloor$ fewer qubits than the design by Thapliyal et al. and has the same qubit cost as the designs by Draper et al. and Trisetyarso et al. Further, In-FT-QCLA2 requires $n + w(n)+\lfloor log(n)\rfloor-1$ fewer qubits than the designs by Takahashi et al. in \cite{Takahashi2008qcla} and \cite{Takahashi2010inlaceQCLA}.  In-FT-QCLA1 achieves its T gate savings with only an order $\mathcal{O}(1)$ increase in qubits compared to the existing work.  

\section{Existing Work}
\label{prior-works-qcla}
Quantum carry lookahead adders (QCLA) have caught the attention of researchers and many have contributed many designs to the literature.  Several designs such as \cite{Babu2013QCLAoutofplace} \cite{Lisa2015QCLAoutofplace} target reversible computing and, therefore, produce significant garbage output.  Other designs such as \cite{Nagamani2016adhocQCLA} present promising designs but they are not generic, prohibiting scaling the designs to alternative input qubit lengths.  Designs that can be implemented on quantum hardware include \cite{Babu2013QCLAoutofplace} and \cite{Draper2006qcla}. In-place and out-of-place QCLAs have been proposed.  We discuss the existing work for in-place and out-of-place QCLAs in separate sections. 

\subsection{Out-of-Place QCLAs}

Existing out-of-place QCLAs that can be implemented on quantum hardware include \cite{Babu2013QCLAoutofplace} \cite{Lisa2015QCLAoutofplace} \cite{Draper2006qcla} \cite{Trisetyarso2009qcla} \cite{Thapliyal2013qcla}.  Table \ref{Comparison-of-OOP-works} summarizes important performance measures for these QCLAs.  T-count, qubit cost, gates used and if the design produces garbage output are shown.  The designs in \cite{Babu2013QCLAoutofplace} and \cite{Lisa2015QCLAoutofplace} produce garbage output and must be made garbageless before use in quantum algorithms.  As a result, the T-count cost is doubled and the qubit cost is increased by at least $n+1$.  The reported T-count and qubit cost reflect the added cost from removing garbage outputs.  The designs in \cite{Draper2006qcla} \cite{Trisetyarso2009qcla} and \cite{Thapliyal2013qcla} have no garbage outputs and can be used as is.  The design in \cite{Lisa2015QCLAoutofplace} offers the lowest T-count in exchange for a $\mathcal{O}(1)$ increase in qubit cost compared to existing works.  The designs in \cite{Draper2006qcla} \cite{Trisetyarso2009qcla} and \cite{Thapliyal2013qcla} achieve the lowest qubit cost with only a modest $\mathcal{O}(1)$ increase in T-count compared to more T-count efficient works such as \cite{Lisa2015QCLAoutofplace}.  The design in \cite{Babu2013QCLAoutofplace} has the highest resource costs of the existing works.  These are all interesting works that offer options with reduced qubit or T gate costs.  However, with advances such as recent T gate efficient Toffoli gate implementations shown in  \cite{Gidney20184TgateToffolibuild}, we have designed quantum QCLAs that have reduced T-count compared to these works.  Further, we can design quantum QCLAs that offer T-count savings yet maintain comparable qubit costs.          

\subsection{In-Place QCLAs}

Existing in-place QCLAs that can be implemented on quantum hardware include \cite{Draper2006qcla} \cite{Takahashi2008qcla} \cite{Trisetyarso2009qcla} \cite{Thapliyal2013qcla} \cite{Cheng2002QCLAinplace} and \cite{Mogensen2019qcla}.  Table \ref{Comparison-of-IP-works} summarizes important performance measures for these QCLAs.  T-count, qubit cost, gates used and if the design produces garbage output are shown.  All the designs have no garbage outputs and can be used as is.  The in-place QCLA in \cite{Cheng2002QCLAinplace} is based on CNOT, Toffoli gates and Multiple Control Toffoli gates.  To decompose a multiple control Toffoli gate into quantum gates, first the multiple control Toffoli gate must be decomposed into Toffoli gates.  One multiple control Toffoli gate translates into $2 \cdot n -3$ Toffoli gates which has a corresponding T-count of $14 \cdot n - 21$ (using the Toffoli gate implementation in \cite{Maslov}).  As a consequence, the design in \cite{Cheng2002QCLAinplace} has a T-count of order $\mathcal{O}(n^3)$ which becomes prohibitively costly for large $n$.  The other existing works have more reasonable T gate costs of order $\mathcal{O}(n)$.  Of these works, the designs by Mogenson have the lowest T-count and can perform their operation with only $3\mathcal{O}(n)$ qubits.  Among the works with T gate costs of order $\mathcal{O}(n)$, the design in \cite{Takahashi2008qcla} is the most costly in terms of T-count and qubit costs.  Prior works such as \cite{Trisetyarso2009qcla} \cite{Thapliyal2013qcla} and \cite{Mogensen2019qcla} present QCLAs that offer low T-count, low qubit cost and no garbage outputs.  However, with the recent T gate efficient Toffoli gate implementations shown in \cite{Gidney20184TgateToffolibuild}, we have designed quantum QCLAs that have reduced T-count compared to these works.  Further, we have designed quantum QCLAs that offer T-count reduction with only a modest $\mathcal{O}(1)$ qubit cost increase compared to existing work.   

\section{Conclusion}

In this work, we propose quantum circuits for carry lookahead addition.  We present proposed designs for in-place QCLAs and out-of-place QCLAs.  We present three designs for in-place QCLAs and three designs for the out-of-place QCLAs.  The in-place FT-QCLA1 (In-FT-QCLA1) and out-of-place FT-QCLA1 (Out-FT-QCLA1) are optimized for T-count.  The in-place FT-QCLA2 (In-FT-QCLA2) and out-of-place FT-QCLA2 (Out-FT-QCLA2) are optimized for qubit cost while providing low T gate cost.  The proposed QCLAs are based on NOT gates, CNOT gates, Toffoli gates, logical AND gates, uncomputation gates as well as a proposed uncomputation gate for near term quantum hardware.  These designs are compared and shown to have reduced T gate and qubit costs compared to the existing work.  We conclude that the proposed in-place QCLAs and out-of-place QCLAs can be used in larger quantum data-path circuits where gate count and/or qubit cost is of concern.  We also conclude that the proposed QCLAs can be used to increase the amount of computation possible on quantum hardware with limited coherence times.

\bibliographystyle{IEEEtran}
\bibliography{references.bib}

\end{document}